%% file: main.tex
\documentclass[conference]{IEEEtran}
\IEEEoverridecommandlockouts
\usepackage{amsmath,amssymb}
\usepackage{graphicx}
\usepackage{braket}
\usepackage{epsfig}
\usepackage{epstopdf}
\usepackage{mdwmath}
\usepackage{xcolor}
\usepackage{varwidth}
\usepackage{framed}
\usepackage{float}
\usepackage{placeins}
\usepackage{afterpage}
\usepackage{ragged2e}
\usepackage{amssymb}
\usepackage{pifont}

\usepackage{blindtext}
\usepackage{enumitem}
\usepackage{xcolor}
\usepackage{eqparbox}
\usepackage{fixltx2e}
\usepackage{multirow}
\usepackage{enumitem,lipsum}

\usepackage[english]{babel}
\newcolumntype{P}[1]{>{\centering\arraybackslash}p{#1}}
\usepackage{graphicx}
\usepackage{amsmath, amssymb}
\usepackage{graphicx}
\usepackage{epsfig}
\usepackage{epstopdf}
\usepackage{array}
\usepackage{graphicx}
\usepackage{subcaption}
\usepackage{hyphenat}

\usepackage{booktabs} % For formal tables
\usepackage{graphicx}
\usepackage{caption}
\usepackage{bbding}
\usepackage{pifont}
\usepackage{wasysym}
\usepackage{amssymb}

\usepackage{framed,url,caption,graphicx}
\usepackage{soul,xspace,tabularx}

\expandafter\def\expandafter\UrlBreaks\expandafter{\UrlBreaks
  \do\a\do\b\do\c\do\d\do\e\do\f\do\g\do\h\do\i\do\j%
  \do\k\do\l\do\m\do\n\do\o\do\p\do\q\do\r\do\s\do\t%
  \do\u\do\v\do\w\do\x\do\y\do\z\do\A\do\B\do\C\do\D%
  \do\E\do\F\do\G\do\H\do\I\do\J\do\K\do\L\do\M\do\N%
  \do\O\do\P\do\Q\do\R\do\S\do\T\do\U\do\V\do\W\do\X%
  \do\Y\do\Z}

\providecommand{\myparab}[1]{\smallskip\noindent\textbf{#1} }

\newcommand{\eat}[1]{}

\usepackage{braket}
\usepackage[utf8]{inputenc}
\usepackage{algorithm2e}
\makeatletter
\renewcommand{\@algocf@capt@plain}{above}% 
\makeatother

\usepackage[utf8]{inputenc}
\usepackage[english]{babel}

\usepackage{lastpage}

\usepackage{indentfirst}
\usepackage[T1]{fontenc}
\usepackage[utf8]{inputenc}
\usepackage{authblk}
\usepackage{graphicx}
\begin{document}

\title{A Quantum Overlay Network for Efficient Entanglement Distribution}
% \author{Paper Number $1570831982$}

\author[1]{Shahrooz Pouryousef\thanks{shahrooz@cs.umass.edu}}
\author[2]{Nitish K. Panigrahy\thanks{nitishkumar.panigrahy@yale.edu}}
\author[1]{Don Towsley\thanks{towsley@cs.umass.edu}}
\affil[1]{College of Information and Computer Sciences, UMass Amherst}
\affil[2]{School of Engineering \& Applied Science, Yale University}

\renewcommand\Authands{ and }

\maketitle
\thispagestyle{plain}
\pagestyle{plain}

\input{abstract}


\input{introduction_new}

\input{preliminaries}

\input{QON_models}

\input{evaluation.tex}

\label{evaluation}

\input{relatedworks-new}

\input{conclusion}
\input{appendix}

\bibliographystyle{plain}
\bibliography{sample-bibliography}

\end{document}

%% file: abstract.tex
\begin{abstract}
Distributing quantum entanglements over long distances is essential for the realization of a global scale quantum Internet. Most of the prior work and proposals assume an on-demand distribution of entanglements which may result in significant network resource under-utilization. In this work, we introduce Quantum Overlay Networks (QONs) for efficient entanglement distribution in quantum networks. When the demand to create end-to-end user entanglements is low, QONs can generate and store maximally entangled Bell pairs (EPR pairs) at specific overlay storage nodes of the network. Later, during peak demands, requests can be served by performing entanglement swaps either over a direct path from the network or over a path using the storage nodes. We solve the link entanglement and storage resource allocation problem in such a QON using a centralized optimization framework. We evaluate the performance of our proposed QON architecture over a wide number of network topologies under various settings using extensive simulation experiments. Our results demonstrate that QONs fare well by a factor of $40\%$ with respect to meeting surge and changing demands compared to traditional non-overlay proposals. QONs also show significant improvement in terms of average entanglement request service delay over non-overlay approaches. 
\end{abstract}

%% file: introduction_new.tex
\section{Introduction}
% \myparab{Quantum Internet}
\vspace{-0.03in}
The vision of a quantum Internet, a global network capable of transmitting quantum information, brings with it the promise of implementing quantum applications such as quantum key distribution (QKD) \cite{bennett2020quantum}, quantum  computation \cite{cirac1999distributed}, quantum sensing \cite{d2001using}, clock synchronisation \cite{komar2014quantum}, quantum-enhanced measurement networks \cite{giovannetti2004quantum}, and many others \cite{kimble2008quantum}. Recent experiments have demonstrated successful quantum key distribution at short distances \cite{peev2009secoqc,wang2014field,stucki2011long} and some are even commercially available \cite{giovannetti2004quantum}.

Executing these applications relies on creating long distance quantum entanglements between end nodes in a quantum entanglement distribution network \cite{lloyd2004infrastructure,dahlberg2019link}. Quantum entanglement is a shared state between two or more quantum bits (qubits) where the quantum state of individual qubits cannot be described independently of the others. Distributing quantum entanglement over long distances first involves creating entanglements, known as link-level entanglements, between adjacent nodes in a quantum network. Finally, end-to-end user entanglements are generated by having each node on the path connecting the end users perform a quantum operation known as an \textit{entanglement swap} on the individual link-level entanglements. These end-to-end user entanglements can then be delivered to respective quantum applications such as QKD for consumption.

As more and more quantum Internet based applications develop in the future, the ever-evolving nature of quantum applications will bring new challenges in managing and delivering quantum entanglement services to end users. Most studies of quantum networks have focused on developing routing protocols for entanglement generation \cite{zhao2021redundant,pant2019routing,chakraborty2020entanglement}, \cite{victora2020purification,zhao2022e2e} and network protocol stack design \cite{lloyd2004infrastructure,dahlberg2019link}. Existing proposals for quantum networks generally assume static conditions and cannot serve end user demands at more than a predefined capacity at any given time.  We envision time varying user traffic demands for end-to-end user entanglements in future quantum networks. Thus, situations may arise when the quantum network cannot generate entanglements at the requested rate. We pose the following research question. {\it What networking infrastructure will allow for handling  peak traffic demands for creating end-to-end user entanglements in a quantum network?} We introduce Quantum Overlay Networks (QONs) as a solution to address this problem.

\begin{figure}
\centering
    \includegraphics[width=8cm]{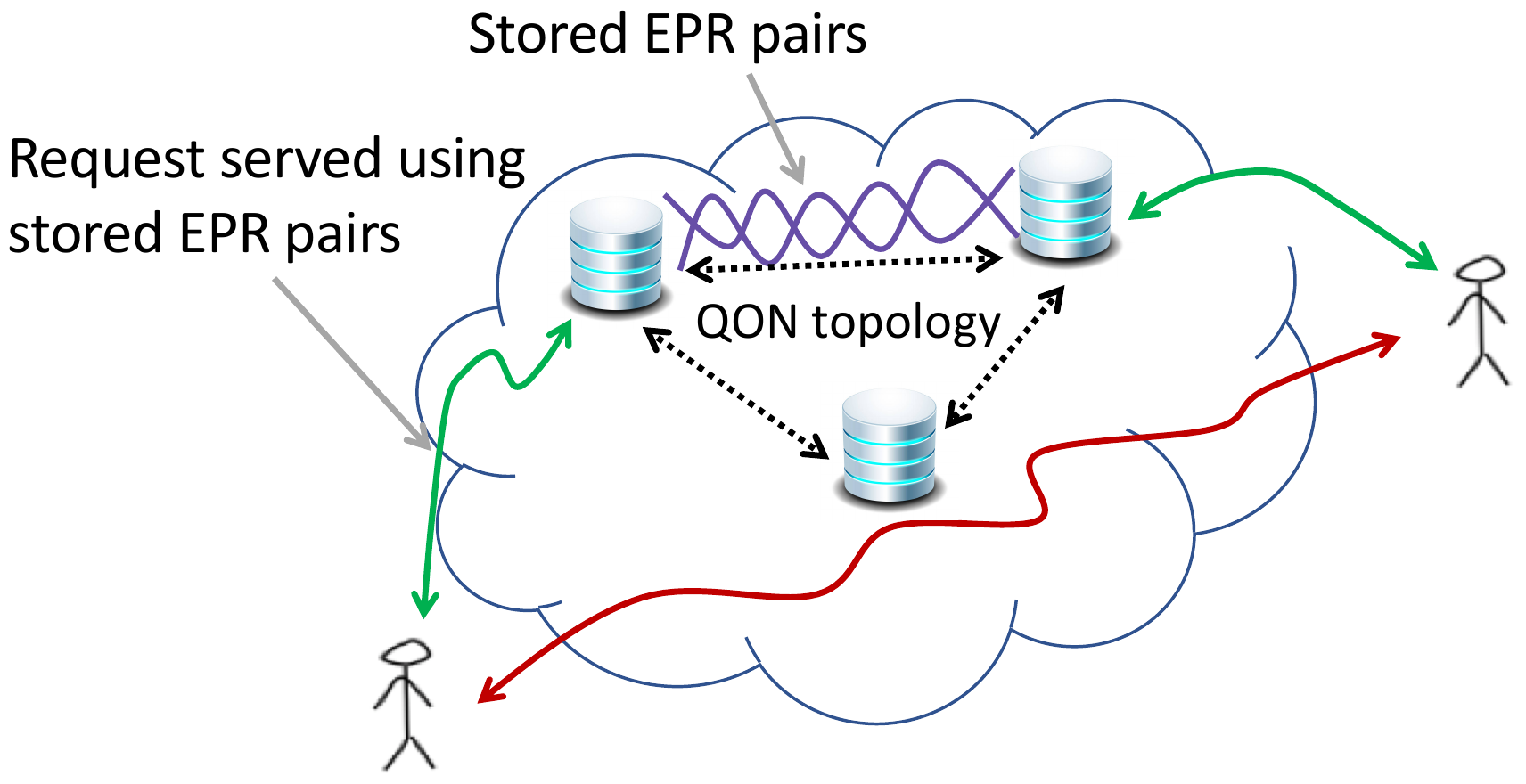}
  \caption{A pictorial depiction of serving user demands in a QON. Users can create end-to-end entanglement using either the entanglements stored at storage nodes (green arrows) or consuming link level entanglements directly from the network (red arrow). \label{fig:QON_overview}}
\end{figure}

We briefly describe the QON architecture. In a QON, we envision that the network infrastructure  provides entanglement storage service at some of the nodes of the network for future usage. The infrastructure may deploy a number of {\em storage nodes} each with a set of long-lived quantum memories as shown in Figure \ref{fig:QON_overview}. Recent advances in quantum memory design suggest that memory lifetimes on the order of $16$ seconds \cite{dudin2013light,longdell2005stopped} and up to one minute \cite{heinze2013stopped,ma2021one} are achievable. Thus it is not unreasonable to expect quantum memory coherence times of a few minutes in 5-10 years.

However, QMs are likely to operate at cryogenic temperatures and may require isolated and costly hardware to operate. Thus, these resources will be limited and it is important to manage them efficiently in a QON. In periods of low demand, a quantum network can dedicate its resources (link-level entanglements) to creating entanglements between the storage nodes in addition to generating entanglement directly between users. During high demands, requests can be served by performing entanglement swaps either over a direct paths connecting end users (red solid arrow) or over paths using the storage nodes (green solid arrows) as shown in Figure \ref{fig:QON_overview}. 

When user demands are dynamic and stochastic, the QON resources (e.g., link-level entanglements and storage capacity) need to be manged carefully. In particular, the following key trade-off should be addressed. {\it How much of the network resources should be used to serve present user demands versus how much of the resources should be allocated for entanglement generation between storage nodes for future usage?} In this work, we precisely answer these questions. We formulate and solve several QON resource allocation problems with different performance objectives. These include serving peak end user demands, minimizing average request service delay, and maximizing weighted entanglement generation rate.

Apart from end-to-end entanglement generation rate, another important network performance metric is the end-to-end entanglement fidelity. Here, entanglement fidelity is a measure of the quality of a served entanglement. Due to noise in quantum channels and quantum operations, fidelity decreases with each quantum swap operation. Quantum applications may set a minimum threshold on end-to-end entanglement fidelity. End-to-end entanglement fidelity can be improved through an entanglement purification quantum operation \cite{Dur1999}. Some end-to-end entanglements are sacrificed in the purification process, which decreases end-to-end entanglement generation rate. In this work, we apply purification on established end-to-end entanglements to satisfy the fidelity requirements of end-user applications.

Our contributions are summarized below.

\begin{itemize}
\item We propose and develop a QON architecture that places nodes with long-lived QMs in a quantum network to efficiently distribute quantum entanglements. To the best of our knowledge, this is the first work that presents a complete design and performance analysis of a quantum overlay network. 

\item We present several optimization formulations for a QON to optimize different performance objectives such as aggregate entanglement generation rate, and average entanglement request service delay.

\item We evaluate the effectiveness of our proposed solution through experiments on both real-world classical networks (ex- ATT, IBM, Abilene, SURFnet) and random networks (power law and Erdos Renyi). Our results confirm that QONs can satisfy peak entanglement generation requests about $40\%$ more than non-overlay approaches. QONs also increase the weighted entanglement generation rate, and significantly reduce the request service delay over non-overlay proposals.
\end{itemize}

The rest of the paper is organized as follows. First, we present some preliminary background information on quantum operations and the system model in Section \ref{sec:Preliminaries}. In Section \ref{sec:formulations}, we explain different models and objectives of our proposed QON architecture. In Section \ref{sec:evaluation}, a thorough performance evaluation of the proposed architecture is conducted and conclusions are drawn in Section \ref{sec:concl}.

%% file: preliminaries.tex
\section{Technical Preliminaries}
\label{sec:Preliminaries}
% \vspace{-0.03in}
In this section, we provide a high level overview of some of the quantum operations and describe the system model that will be used throughout the paper. 

\subsection{Quantum Background}
% \vspace{-0.02in}
We now briefly mention some of the quantum operations that are relevant to this work.

\myparab{Quantum States:}
A quantum bit (qubit) is fundamentally different than a classical bit. A classical bit can either be in state $0$ or $1$ while a qubit can be in a superposition state of state $0$ and $1.$ Typically a qubit is represented as $|\alpha \rangle= \alpha_1|{0}\rangle +\alpha_2|{1}\rangle$. Similar to one qubit system, a two qubit quantum system can be in a superposition of $|00 \rangle,
|01 \rangle, |10\rangle$, $|11 \rangle$ states and represented as $|\tilde{\alpha} \rangle= \tilde{\alpha}_{1}|{00}\rangle +\tilde{\alpha}_{2}|{01}\rangle+\tilde{\alpha}_{3}|{10}\rangle +\tilde{\alpha}_{4}|{11}\rangle$. While there exists many possible physical realization of a qubit, photonic qubits are most likely to be used for quantum communication and information transfer. In practice, these qubits can be sent through optical fiber based quantum channels using  polarization, time-bin, or absence and presence of a photon based encodings \cite{mattle1996dense}.

\myparab{Quantum Entanglement:}
An entangled state is a special type of multi-qubit quantum state which can not be written as the product of its individual component states. The measurement outcomes of an entangled state are correlated. A two qubit maximally entangled state shared between two parties $A$ and $B$ can be in one of the following four forms, also known as the {\it Bell pairs}: $|\psi_{AB}^{\pm}\rangle=\frac{|{0_A0_B}\rangle\pm|{1_A1_B}\rangle}{\sqrt[]{2}}$ and $|\phi_{AB}^{\pm}\rangle=\frac{|{0_A1_B}\rangle\pm|{1_A0_B}\rangle}{\sqrt[]{2}}$. We refer to Bell pair $\ket{\psi_{AB}^+}$ as an {\it EPR pair}. 

\myparab{Entanglement Swapping:}
Suppose two parties $A$ and $B$ share an EPR pair $\ket{\psi_{AB}^+}$. Also, assume that $B$ shares another pair $\ket{\psi_{BC}^+}$ with $C$. Then $B$ can create an EPR pair $\ket{\psi_{AC}^+}$ between $A$ and $C$ by performing a bell state measurement followed by classical communication exchange and a correction. This operation is known as \textit{entanglement swapping}. The process can be repeated in a nested manner to create EPR pairs between distant parties.

\myparab{Fidelity and Entanglement Purification:}
Fidelity is a widely used metric to quantify the quality of an entanglement. Due to noisy quantum channels, interaction with environment and imperfect quantum operations, the quality of an EPR pair typically decreases from its initial state. However, quantum applications may have a minimum threshold on the fidelity ($F^{th}$) of EPR pairs for subsequent usage. Thus, low quality EPR pairs delivered to the application may not be suitable for consumption. A quantum operation known as {\it Entanglement Purification} can solve this issue. Entanglement purification typically consumes two low-fidelity base EPR pairs and create one EPR pair with higher fidelity. Each purification step is probabilistic and both base pairs are lost in case of failure. This operation can be repeated in a nested manner by again purifying  two newly purified EPR pair until a target fidelity is reached. These protocols are also known as {\it recurrence based purification protocols}. We refer to a purified EPR pair with fidelity greater than target fidelity as a high quality EPR pair.

Let $F$ denote the fidelity of base EPR pairs that participate in purification. The average number $g(F, F^{th})$ of base EPR pairs needed to create one high quality EPR pair under a recurrence based purification protocol is given by \cite{Dur1999},
\begin{align}\label{eq:purification_avg}
    g(F, F^{th}) = \prod\limits_{k=1}^{k_{max}}\frac{2}{p_k},
\end{align}
\noindent where $k_{max}$ denotes the number of successful purification steps needed to achieve an output fidelity of $F^{th}$ starting from an input fidelity $F$ and $p_k$ denotes the success probability of $k^{th}$ purification step.

\myparab{Quantum Memories:}
QMs can store EPR pairs for future usage. To realize long-distance entanglement storage, the photonic qubits need to be stored at intermediate nodes of a quantum network with high efficiencies. In a QM, photonic quantum states are mapped onto individual matter (non-photonic) qubits. Physical implementations of QMs can be grouped into the following five platforms: rare-earth ion-doped solids, diamond color centers, crystalline solids, alkali metal vapours, and molecules. We refer interested readers to \cite{heshami2016quantum,sangouard2011quantum} for a more elaborate discussion on various proposed implementations of QMs.

\subsection{Basic Components of QON}

In this section, we describe the basic components of QON. Table \ref{table:notaions} shows the notations used in this and next sections.

\begin{table}
    \setlength{\tabcolsep}{3.5pt}
    \begin{tabularx}{\columnwidth}{c X }
      \toprule
      {$c(u,v)$} & {Capacity of link $(u,v)$ in EPRs/sec}   \\ 
\\
$K$ & Set of user pairs\\
        \\
{Virtual link} & A newly added link to the graph that connects a storage pair directly \\ 
\\
         {$P_{N}^{k}$}  & {Set of all  paths for user pair $k$ in $G$}  \\
        % \\
        % {$P_{R}^{j}$}  & {Set of all paths between storage pair $j$}  \\
        \\
        {$P^{k}_{S}$}  & {Set of all paths for user pair $k$ that use at least one virtual link in $\tilde{G}$}  \\
        \\
          {$F^{k}_t$}  & Fidelity threshold of user pair $k$ at time interval $t$  \\
        \\
        $g(F_p,F^{th})$ & Avg. no. of base EPR pairs needed for purification on path $p$ to achieve fidelity threshold $F^{th}$\\
        \\
        % $f(F_p,n)$ & Implements the purification scheme and  purify path $p$ with basic fidelity $F_p$ using $n$ EPR pairs\\
        % \\
        $S$ & Set of nodes that have storing capability\\
        \\
        $J$ & Set of storage pairs \\
        \\
    {$D_{t}^{k}$}  & {Rate of demand for request pair $k$  at time interval $t$}   \\
    \\
    {$B_s$}  & {Capacity of storage server $s$}   \\
          
\\

         {$F_{p}$}  & {Basic fidelity of path $p$ } 
                    \\ 
\\
$w_{p,t}^{k}$ & Rate of entanglement generation for pair $k$ on path $p$ at time interval $t$\\
\\

        $u^j_{p,t}$  & Avg. no. of EPR pairs at storage $j$ at the beginning of time interval $t$ that are generated using path $p$\\
        \\
        % $n^k_{p,t}$& Avg. no. of EPR pairs used on path $p$ for user pair $k$ for purification to generate one high-quality EPR pair
        % \\
        % \\
        $\Delta$& Duration of one time interval in sec
\\  
\\
$h$& EPR pair lifetime (e.g., no. of time intervals) at storage servers
\\ 
\hline
        
      \bottomrule
    \end{tabularx}
    \caption{\label{table:notaions} Notations used in this paper.}
\end{table}

\myparab{Network:} We consider a quantum network represented by a graph $G = (V, E)$, where $V$ is the set of nodes and $E$ is the set of physical communication links. We define $c(u,v)$ as the capacity of edge $(u,v)$, which denotes the average EPR pair generation rate between adjacent nodes $u$ and $v$. We assume a subset of the nodes ($S\subset V$) in the network have storage capability and pairs of them can store EPR pairs for satisfying future requests. Let $J$ denote the set of storage pairs in the network. The storage nodes can also act as normal nodes meaning that they can create EPR pairs with neighboring nodes and perform entanglement swapping. Each  storage server can store a limited number of EPR pairs in its memory for future usage. The capacity of each storage server $s \in S$ is identified by $B_s$. 

\myparab{Workload:} We assume the time to be discretized and consists of a set of $T$ discrete time intervals. We denote $K$ to be the set of source-destination pairs of users that generate entanglement requests at each time interval. Let $D^{k}_{t}$ denote the entanglement generation request rate from source-destination pair $k$ at time interval $t$. We assume there is a central controller in the network that receives these requests and orchestrates resources at different time intervals.  Each request from user pair $k$ has an application-level target fidelity threshold indicated by $F^k_t$.

\begin{figure}

  \begin{subfigure}{.24\textwidth}
    \includegraphics[width=4.1cm]{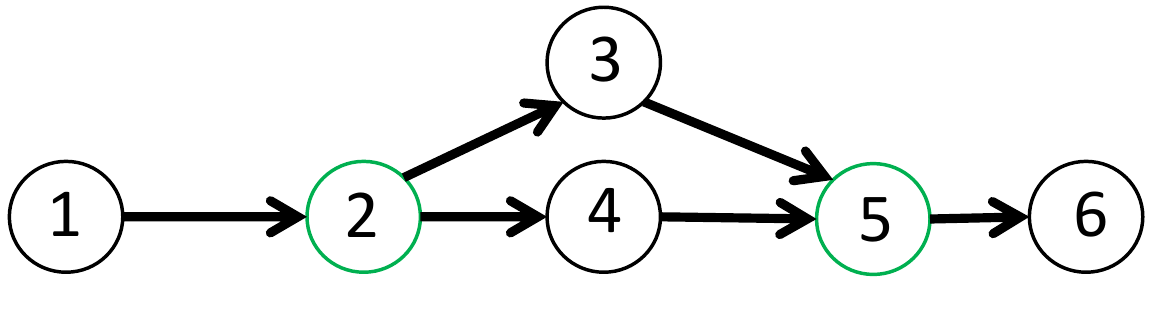}
    % \vspace{-0.1in}
  \caption{Original graph \label{fig:virtual_link-original_graph}}
  \end{subfigure}
  \begin{subfigure}{.24\textwidth}
    \includegraphics[width=4.0cm]{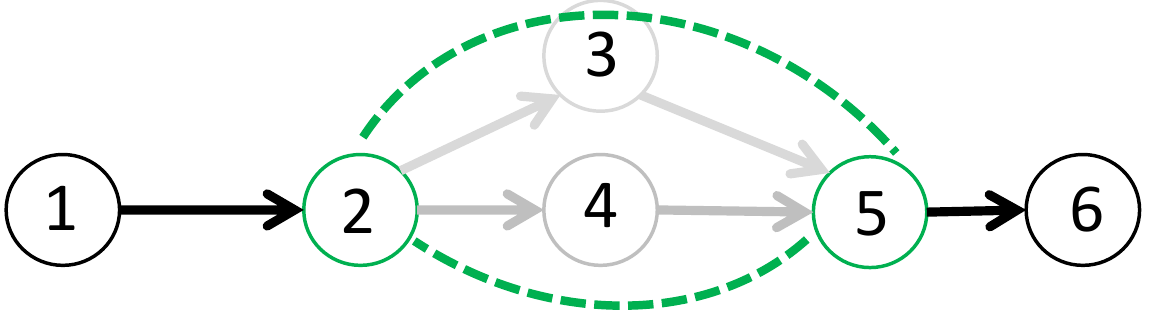}
    % \vspace{-0.06in}
    \caption{Adding one virtual link for each path between storage servers \label{fig:virtual_link_modified_graph}}
  \end{subfigure}
\vspace{-0.09in}
  \caption{Adding parallel virtual links between storage pair $(2,5)$ to construct virtual paths for user pair $(1,6)$.  \label{fig:virtual_links}}
\end{figure}

\myparab{Paths and Virtual Links:} We assume that each user pair $k= (src,dst)$ has a set of predefined paths between them. Same holds true for storage pairs $j = (j_1, j_2).$  We denote the set of paths that connect two user (storage) pairs through the network as $P^{k}_{N}$ ($P^{j}_{N}$). We now add a parallel link between a storage pair for each path that connects them and refer to these new links as {\em virtual} links\footnote{The notion of a single virtual link was first introduced in \cite{Schoute2016}.} (Figure \ref{fig:virtual_link_modified_graph}). Let $\tilde{G} = (V, \tilde{E})$ denote the virtual graph created by adding virtual links between storage node pairs. We define a virtual path to be a path that uses at least one virtual link. For each parallel virtual link, one needs to know the corresponding network path that created it. This will be helpful in tracking the fidelities associated with the EPR pairs stored at the corresponding storage servers. Hence, we associate an identifier for each virtual path. 

For example, in Figure \ref{fig:virtual_links}, consider the virtual path $p_1 = \{1,2,5,6\}$ that was created using network sub-path $s_1 = \{2, 3, 5\}.$ Consider another virtual path $p_2 = \{1,2,5,6\}$ that was created using network sub-path $s_2 = \{2, 4, 5\}.$ We treat $p_1$ and $p_2$ differently and do not count them as duplicates. Let $f(\cdot)$ be a function that outputs the complete path corresponding to a virtual path. In our example: $f(p_1) = \{1,2,3, 5,6\}$ and $f(p_2) = \{1,2,4, 5,6\}$. Let $P^{k}_{S}$ ($P^{j}_{S}$) denote the set of virtual paths between user (storage) pair $k$ ($j$) in $\tilde{G}$.

\myparab{Fidelity of stored EPR pairs:} We need to keep track of the fidelity of stored EPR pairs at each pair of storage servers. Since each storage pair $j\in J$ may use different paths to generate EPR pairs, the fidelity of the stored EPR pairs at each storage pair can be different. For each storage pair and for each path between them, we define a variable $u^j_{p,t}$ to identify the number of EPR pairs generated over path $p$ that are stored at storage pair $j$ at time $t$. All EPR pairs generated over path $p$ have the same basic fidelity. We assume that entanglement purification is always performed by the end-users and storage nodes do not perform any purification.

\myparab{Storage Lifetimes:} We assume the qubits of an EPR pair decohere after being stored for $h$ time intervals and, if not used are thrown out. In Section \ref{sec:formulations}, we consider formulations for resource allocation in a QON for different values of $h$. In particular, we are interested in two extremes: (i) $h = 1$ (unit storage lifetimes) (ii) $h \ge |T|$ (infinite storage lifetimes).

%% file: QON_models.tex
\section{QON: Models and Objectives}
\label{sec:formulations}
% \vspace{-0.06in}
We now introduce the related models and objectives for resource allocation in QONs. 

% \vspace{-0.08in}
\subsection{Handling demand spikes}\label{sub:hds}
\vspace{-0.08in}
One of the goals of this work is to evaluate the robustness of QON in handling sudden demand fluctuations for generating end-user entanglements. Our goal is to serve the demands of all user pairs during each time interval while satisfying their fidelity requirements. We first consider the case that EPR pairs can stay in storage forever and can be used in any time in the future, $h\geq|T|$. We then consider the case that EPR pairs are valid for only one time interval, $h=1$. The third case where EPR pairs can be stored at storage servers for multiple time intervals, $1<h<|T|$, is relegated to Appendix \ref{sec:multi_time_interval_case}.  

\subsubsection{$h \geq|T|$ case}  We now consider the case where EPR pairs can be stored at storage servers forever. Our decision variables $w_{p,t}^{k}$ denote the rate of entanglement generation for user pair $k$ using path $p$ during time interval $t$. Below, we cast the demand satisfaction problem as a constraint feasibility problem and explain its constraints. 

\begin{align} 
\max_{w^k_{p,t}}  & \quad 1 
\nonumber
\\
\text{subject to}\quad
& \quad \forall{t \in T:} \nonumber \\
\quad & u_{p_s,t}^j =u_{p_s,t-1}^{j}  \nonumber\\
\quad  & -\sum_{\substack{k \in \{K\cup \{J-j\}\}\nonumber\\
{p \in P_{S}^{k}}|p_s \subset f(p)}}
        w_{p,t-1}^{k} g(F_p,F^k_{t-1})\Delta \\
\label{cons:inventory_evolution}
\quad & \quad \quad \quad + w_{p_s,t-1}^{j} \Delta  \quad \quad {j\in J}, {p_s \in P^j_{N}\cup P^j_S} \\
\nonumber \\
& \sum_{\substack{k \in \{K\cup \{J-j\}\}\nonumber\\
{p \in P_{S}^{k}}|p_s \subset f(p)}}
w_{p,t}^{k} g(F_p,F^k_{t})\Delta \leq u_{p_s,t}^{j}
\label{cons:inventory_usage}
\nonumber \\
\quad & \quad \quad \quad  \quad \quad \quad {j\in J}, {p_s \in P^j_{N}\cup P^j_S}\\
\nonumber \\
\label{cons:demand_constraint}
\quad & \sum_{ p \in P^{k}_{N}\cup P^k_S}w_{p,t}^{k} =D^{k}_{t}   \quad  {k\in K}\\
\nonumber \\
\quad & \sum_{\substack{k \in \{K \cup J\}\nonumber\\
{p \in P^k_{N}\cup P^k_S}|(u,v) \in p}} w^{k}_{p,t}  g(F_p,F^k_t)  \leq c(u,v)
\label{cons:edge_constraint}
\nonumber \\
\quad & \quad \quad \quad  \quad \quad \quad \quad \quad \quad  \quad \quad \quad {(u,v)\in E}\\
% \quad & \sum_{p_n \in P^{k}_{N}}w_{p_n,t}^{k}  \geq0 \quad \forall{t} \& \forall{k\in J}\\
\quad & \sum_{\substack{s_2 \in S\\
{j = (s,s_2) \in J}\\ p_s \in P^j_{N} \cup P^j_S}}u_{p_s,t}^{j} \leq B_s \quad {{s\in S}} 
\label{cons:storage_capacity_constraint}
\\
\quad & w_{p,t}^{k}\geq0  \quad  {k \in \{K \cup J\}}, {p\in P^{k}_{N} \cup P^k_S}
\label{cons:variable_constraint}
\\
\quad & u^{j}_{p_s,t}\geq0   \quad  \forall{j \in J}, \forall{p_s \in P^j_{N} \cup P^j_S}
\label{cons:storage_variable_constraint}
\end{align}

Here, \eqref{cons:inventory_evolution} captures the evolution of each storage pair $j$. 
$u^{j}_{p_s,t}$ is equal to $u^{j}_{p_s,t-1}$ minus the average number of EPR pairs served from it to satisfy user demands and end-to-end purification in the previous time interval, plus the average number of EPR pairs stored in it using path $p_s$ in the previous time interval. We repeat this constraint for all storage pairs and for all paths between them across all time intervals. Function $f$ returns a complete path $\tilde{p}_s$ corresponding to a virtual path $p$ as explained in Section \ref{sec:Preliminaries}.

Constraint \eqref{cons:inventory_usage} ensures that the average number of purified EPR pairs that are served from one storage pair during time interval $t$ should be less than or equal to what has been available in it at the beginning of the interval. Similar to the previous constraint, we repeat this constraint for all storage pairs and for each path between them that is being used for entanglement generation. For a storage pair j and a given path $p_s$ between them, the condition on the summation term in this constraint captures all paths (end-user pair paths + other inter-storage pair paths) that have $p_s$ as their sub-path.

Constraint \eqref{cons:demand_constraint} ensures that all the demands for each user pair should be satisfied. Constraint \eqref{cons:edge_constraint} ensures that the average number of EPR pairs served or stored using an edge should not be more than its capacity. Constraint \eqref{cons:storage_capacity_constraint} enforces that the storage capacity of each storage node must not be violated. Note that, we have set the value of $u^j_{p,t}$ for $t=0$ to zero, i.e. no EPR pairs are stored at the storage servers at the beginning of first time interval\footnote{Even though here we focus on the formulation for finite $|T|$, one can easily modify our formulation for the case when $T$ is periodic by setting $u^j_{p,0}=u^j_{p,|T|}$}. In addition, the value of $w^k_{p,t}$ for $t=0$ and for any user pair $k$ with $p \in P^{k}_{S}$ is set to zero. This means that we can not serve any EPR pair from the storage servers at the first time interval since nothing has been stored at the storage servers yet. The function $g(F_p,F^k_t)$ in constraints \eqref{cons:inventory_evolution}, \eqref{cons:inventory_usage}, and \eqref{cons:edge_constraint} returns the average number of base EPR pairs with fidelity $F_p$ that needs to be sacrificed in order to create one high quality EPR pair of fidelity at least $F^k_t$. The function $g(F_p,F^k_t)$ can be computed using Equation \eqref{eq:purification_avg}.

\subsubsection{$h=1$ case}\label{subsubh1} The problem formulation for the case when stored EPR pairs have unit storage lifetimes is similar to the infinite lifetime scenario except for the constraint that tracks the evolution of average number of stored EPR pairs across storage node pairs. The formulation is as follows:
\begin{align} 
\max_{w^k_{p,t}}  & \quad 1
%\label{optimization:max_average_fidelity}
\nonumber
\\
\text{subject to}\quad
& \quad \forall{t \in T:} 
\nonumber \\
\quad & u_{p_s,t}^j =w_{p_s,t-1}^{j}\Delta - \nonumber\\
\quad & \sum_{\substack{k \in \{K\cup \{J-j\}\} \nonumber\\
{p \in P_{S}^{k}}|p_s \subset f(p)}}
        w_{p,t-1}^{k} g(F_p,F^k_{t-1})\Delta 
\label{cons:inventory_evolution_one_time_slot}\\
\quad & \quad \quad \quad  \quad \quad \quad {j\in J}\& {p_s \in P^j_{N} \cup P^j_S}\\
\quad & \text{and constraints}  \quad 
\nonumber
\eqref{cons:inventory_usage}, 
\eqref{cons:demand_constraint}, \eqref{cons:edge_constraint}, \eqref{cons:storage_capacity_constraint}, \eqref{cons:variable_constraint}, \eqref{cons:storage_variable_constraint}
\end{align} 

Note that we do not have the variable $u^j_{p_s,t-1}$ in constraint \eqref{cons:inventory_evolution_one_time_slot} anymore.

\subsection{Maximizing Weighted Entanglement Generation Rate}
{\it Entanglement Generation Rate} (EGR) provided by a quantum network is another important performance metric and a subject of great interest in recent quantum network proposals. We now investigate a scenario where a population of end-user pairs get served by the quantum network, possibly at different rates. In a quantum network, some end user pairs may have different priority or reward when they get served. Thus, the user pairs can compete for access to QON resources. Let $\alpha^k_t$ denote the weight associated with user pair $k$ at time interval $t$. Our goal is to solve the resource allocation problem in a QON that maximizes the aggregate weighted EGR across all user pairs and all time intervals. The optimization formulation for $h \ge |T|$ is presented below.

\begin{align} 
\max_{w^k_{p,t}}  & \quad \frac{1}{|T|}\sum_{t\in T}\sum_{k\in K}\sum_{p\in P^k_{N} \cup P^k_S}\alpha^k_t * w^k_{p,t} 
%\label{optimization:max_average_fidelity2}
\\
\text{subject to}\quad
\nonumber
& \quad \forall{t \in T:}\\
\quad & \text{constraints}  \quad 
\nonumber
\eqref{cons:inventory_evolution}, \eqref{cons:inventory_usage}, \eqref{cons:edge_constraint}, \eqref{cons:storage_capacity_constraint}, \eqref{cons:variable_constraint}, \eqref{cons:storage_variable_constraint}
\end{align}

Note that, the constraints for this formulation are similar to the formulation mentioned in Section \ref{sub:hds} except that we do not have a notion of end-user demands. Instead we are interested in maximizing aggregate weighted EGR. The formulation for $h = 1$ scenario can be described similar to Section \ref{subsubh1}.

\subsection{Minimizing request service delay}
\label{sec:moving_to_edge}
Similar to the classical overlays, QONs can also be used to minimize the aggregate request service delay while satisfying user request requirements. The key idea is that the EPR pairs at the storage nodes are readily available for consumption. Thus, when requests are served using EPR pairs from storage nodes, one needs to perform lesser number of entanglement swap operations compared to the case when they are served using the network which in turn decreases the end-to-end request service delay. We present the formulation as follows.

\begin{align} 
\min_{w^k_{p,t}}  & \quad \sum_{t\in T}\sum_{k\in K}\sum_{p\in P^k_{N\cup S}}w^k_{p,t} (|p|-1)
%\label{optimization:max_average_fidelity3}
\\
\text{subject to}\quad
& \quad \forall{t \in T:} 
\nonumber \\
\quad & \text{Constraints}  \quad 
\nonumber
\eqref{cons:inventory_evolution}, \eqref{cons:inventory_usage}, \eqref{cons:demand_constraint}, \eqref{cons:edge_constraint}, \eqref{cons:storage_capacity_constraint}, \eqref{cons:variable_constraint}, \eqref{cons:storage_variable_constraint}
\end{align} 
\noindent Here, $|p|$ is the length of path $p$. When $p \in P^k_S,$ the path length greatly reduces due to the notion of virtual links.

\subsection{Complexity Analysis}

The computational complexity of our problem formulation can be analyzed as follows. First of all, the decision variables $w^k_{p,t}$, are continuous. The constraints and objectives mentioned in all previous formulations are linear. Thus all of our optimization formulations are linear programs. Assume $|K|$, $|P|$, and $|E|$ represents the number of user pairs, the number of paths used for each user pair, and number of edges in the network respectively. Our optimization problem has at most $|K|*|P|*|T|$ number of variables and the number of constraints are $\mathcal{O}(|T|*|J|*|P|*|K|)^3+\mathcal{O}(|T|*(|K|+|E|)*|P|))$.

%% file: evaluation.tex
\section{Performance Evaluation}
\label{sec:evaluation}

\begin{figure*}[t]
\centering
\begin{subfigure}{.24\textwidth}
    \centering
    \includegraphics[width=.9\linewidth]{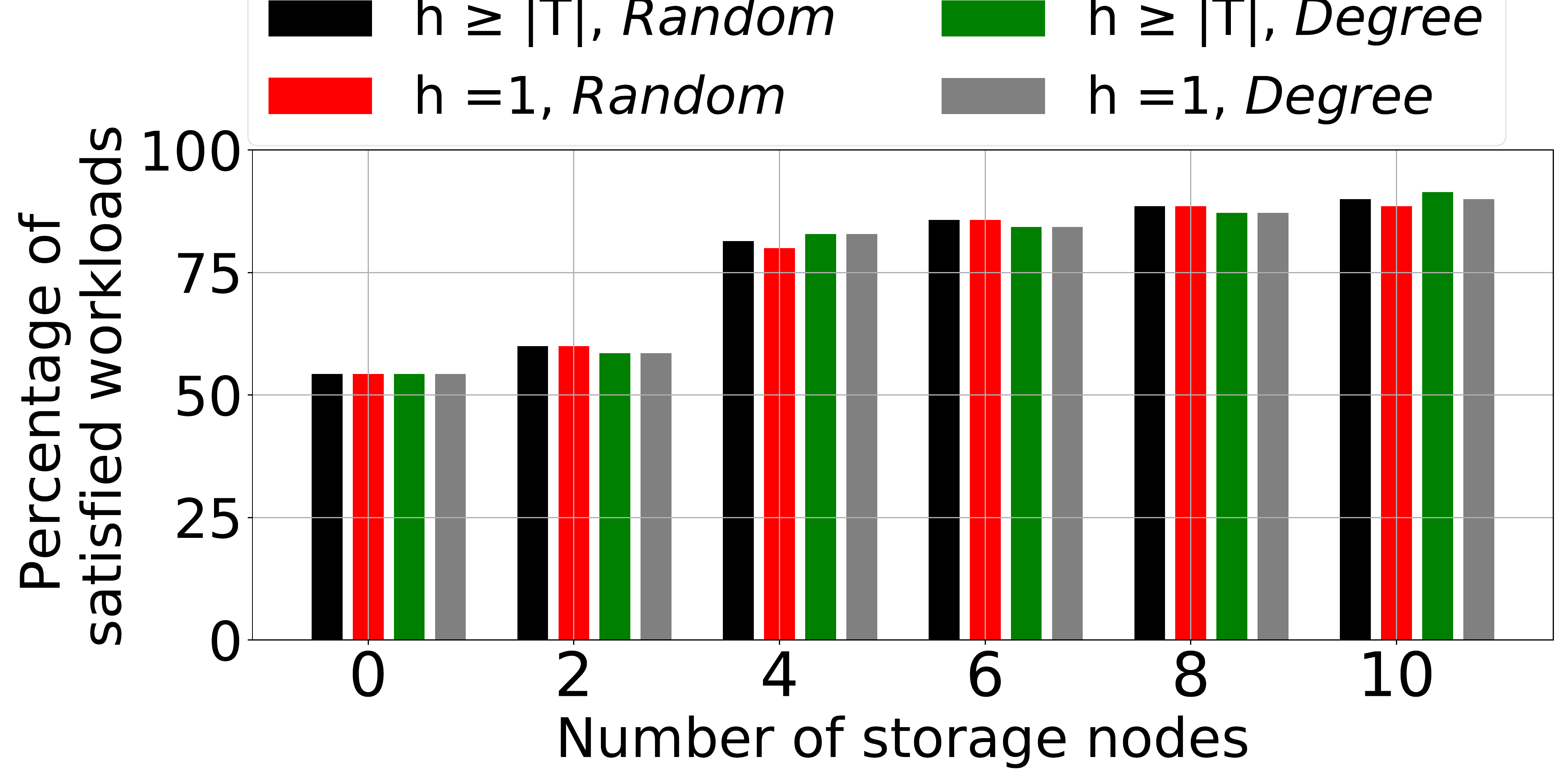}  
     \vspace{-0.06in}
    \caption{ATT}
    \label{fig:satified_in_att}
\end{subfigure}
\begin{subfigure}{.24\textwidth}
    \centering
    \includegraphics[width=.9\linewidth]{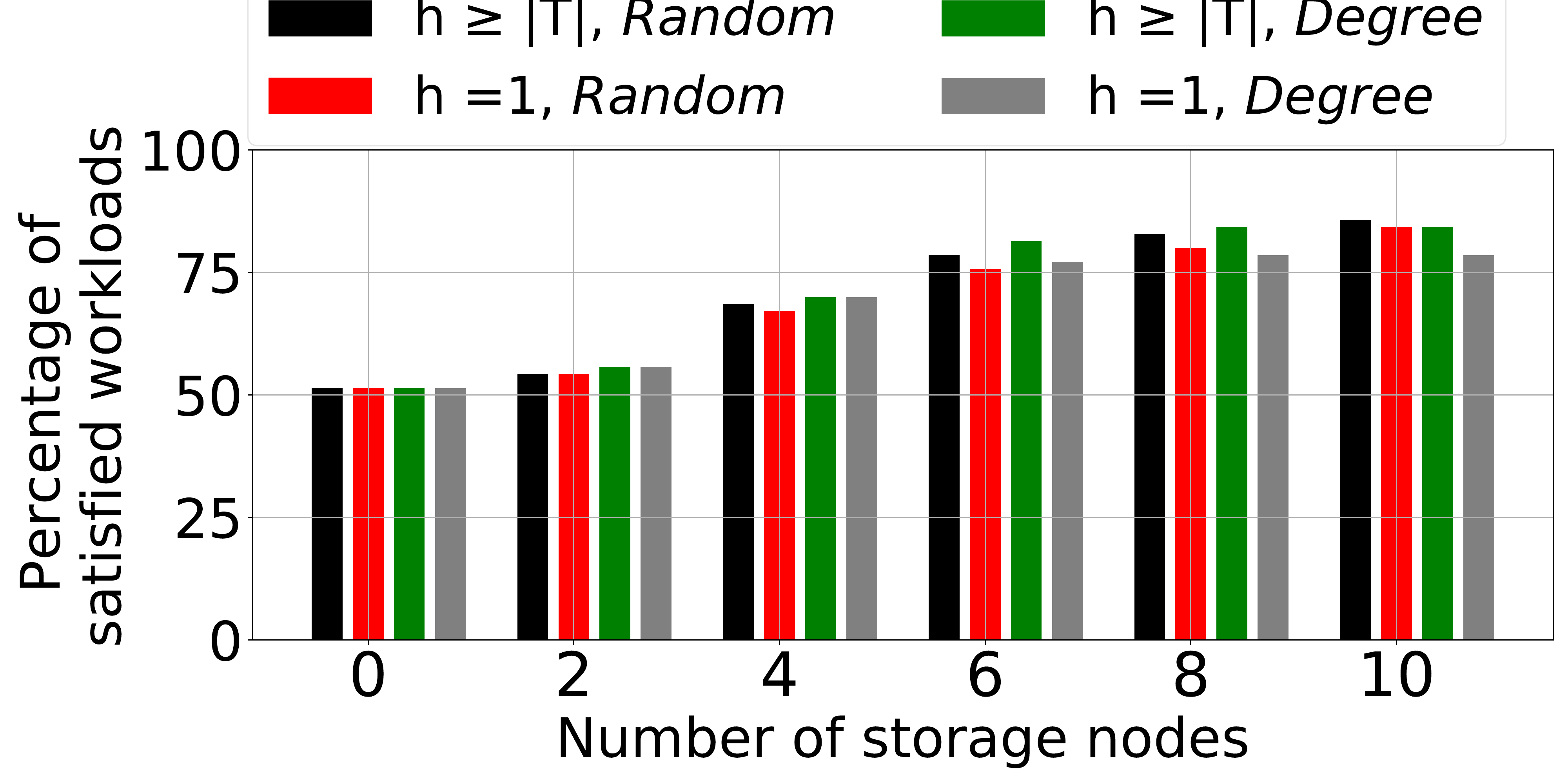} 
    \vspace{-0.06in}
    \caption{Abilene}
    \label{fig:satified_in_abilene}
\end{subfigure}
\begin{subfigure}{.24\textwidth}
    \centering
    \includegraphics[width=.9\linewidth]{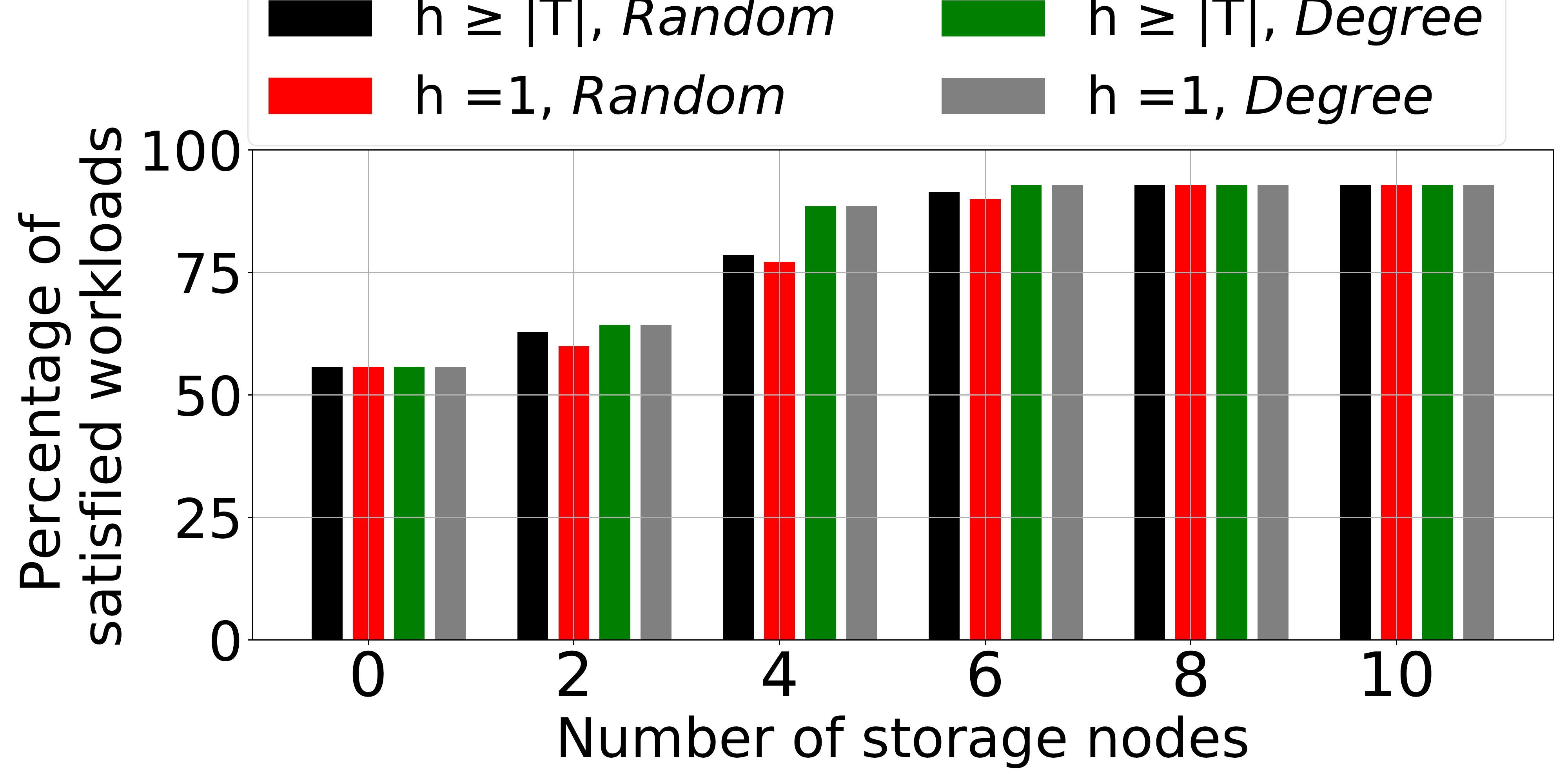}
    \vspace{-0.06in}
    \caption{IBM}
    \label{fig:satified_in_surfnet}
\end{subfigure}
\begin{subfigure}{.24\textwidth}
    \centering
    \includegraphics[width=.9\linewidth]{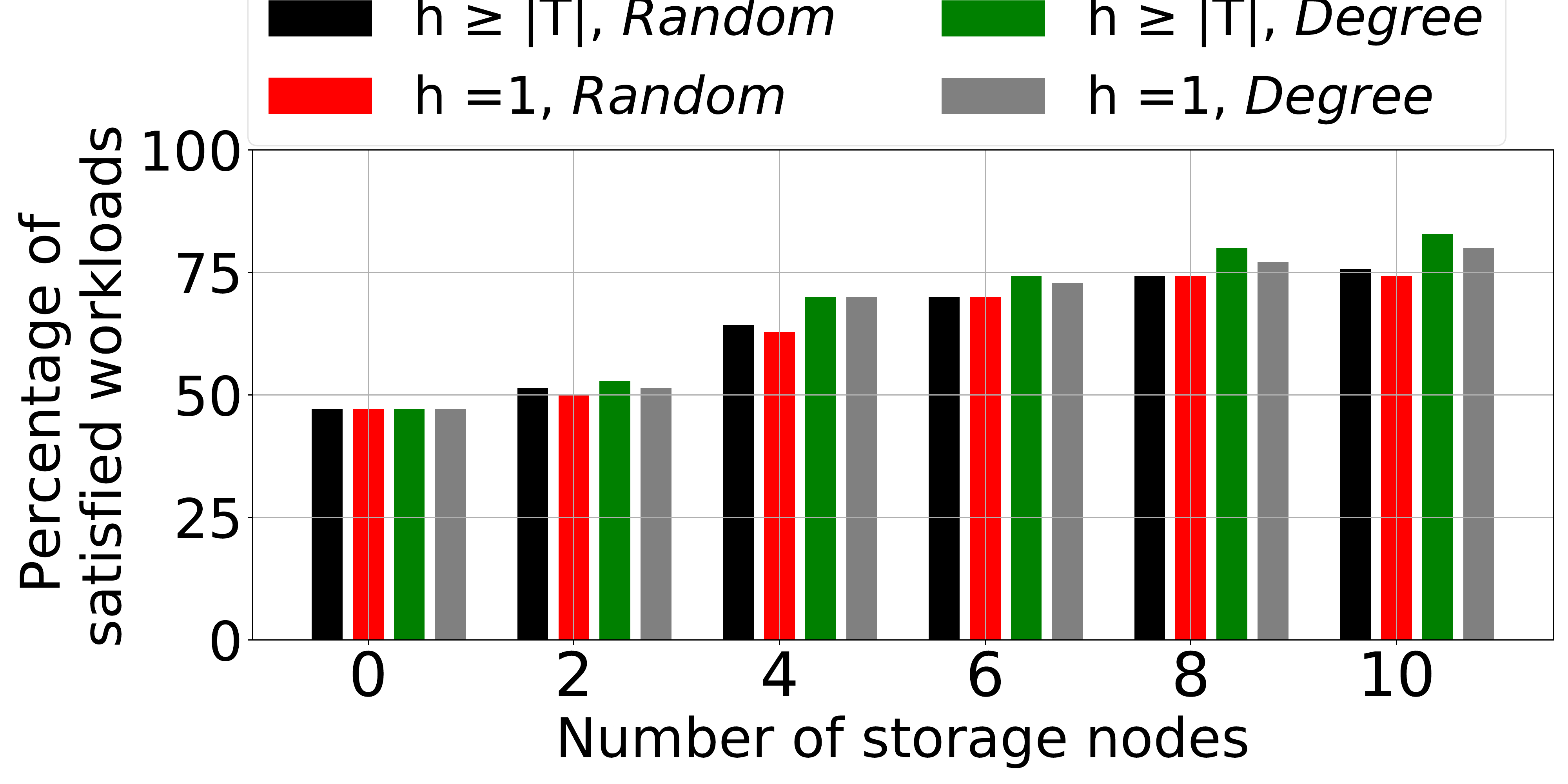} 
    \vspace{-0.06in}
    \caption{SURFnet}
    \label{fig:satified_in_surfnet}
\end{subfigure}
\begin{subfigure}{.24\textwidth}
    \centering
    \includegraphics[width=.9\linewidth]{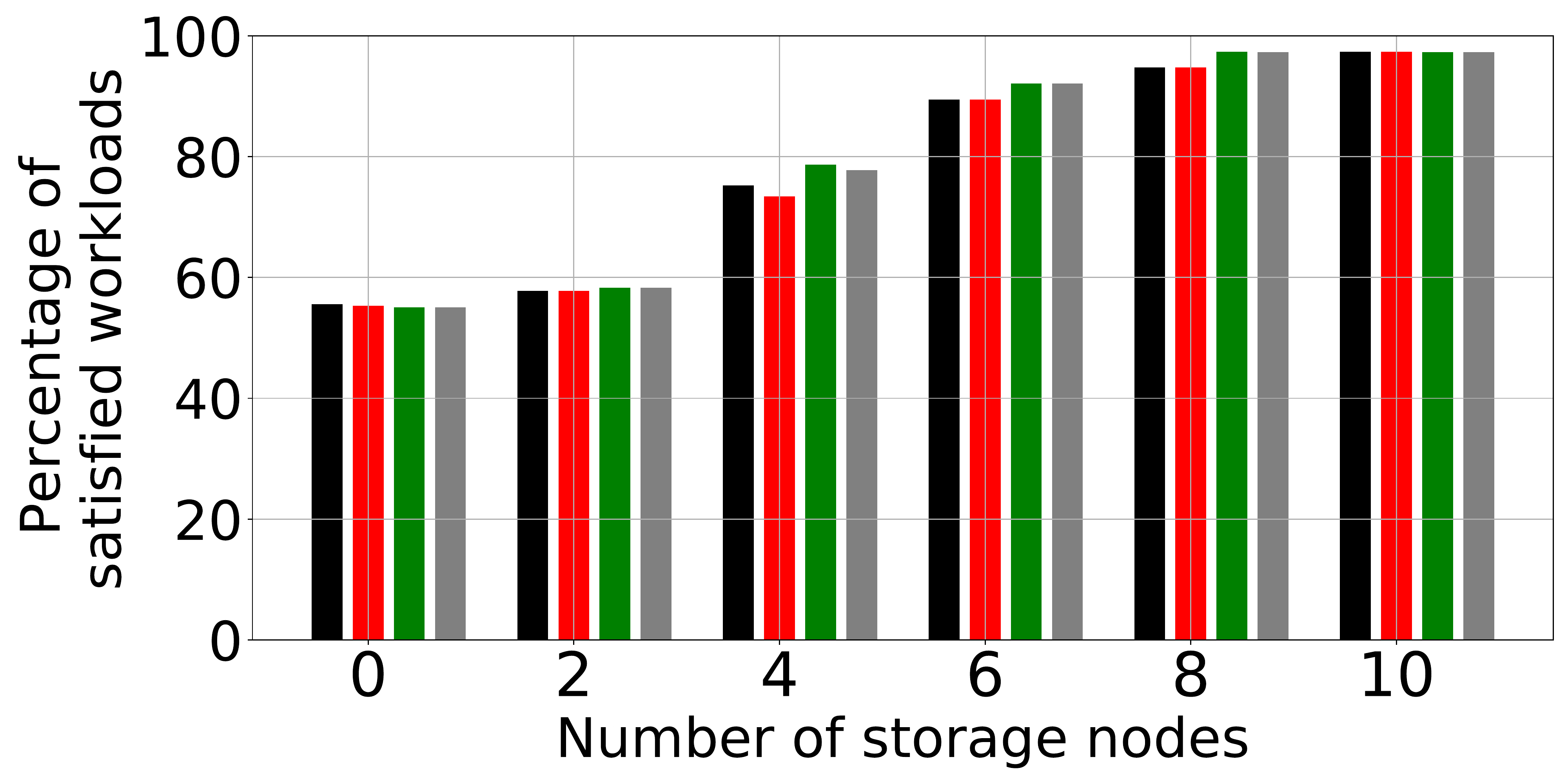}
    \vspace{-0.06in}
    \caption{$G(50,0.1)$}
    \label{fig:satified_in_random1}
\end{subfigure}
\begin{subfigure}{.24\textwidth}
    \centering
    \includegraphics[width=.9\linewidth]{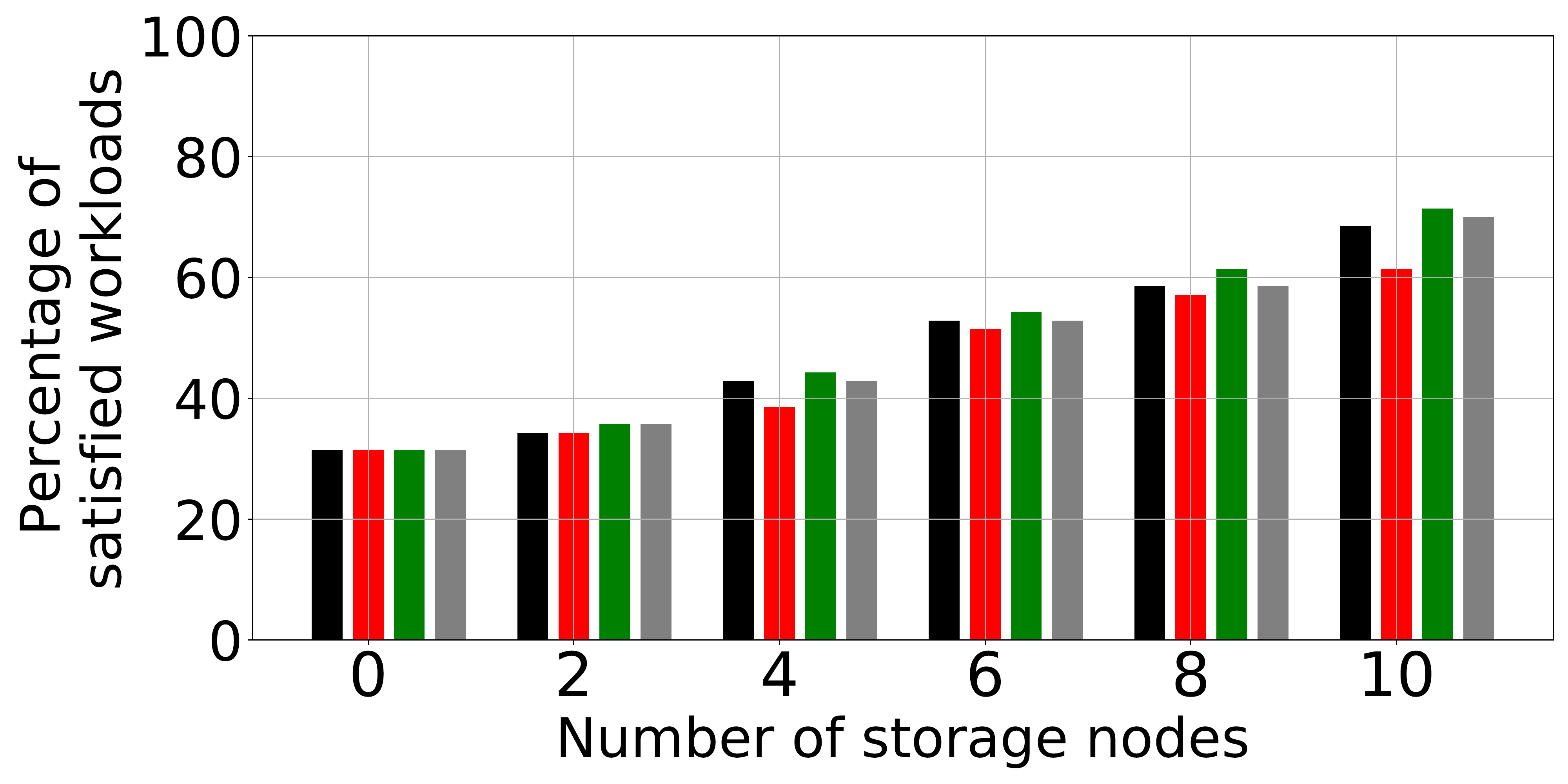}  
    \vspace{-0.06in}
    \caption{$G(50,0.05)$}
    \label{fig:satified_in_random2}
\end{subfigure}
\begin{subfigure}{.24\textwidth}
    \centering
    \includegraphics[width=.9\linewidth]{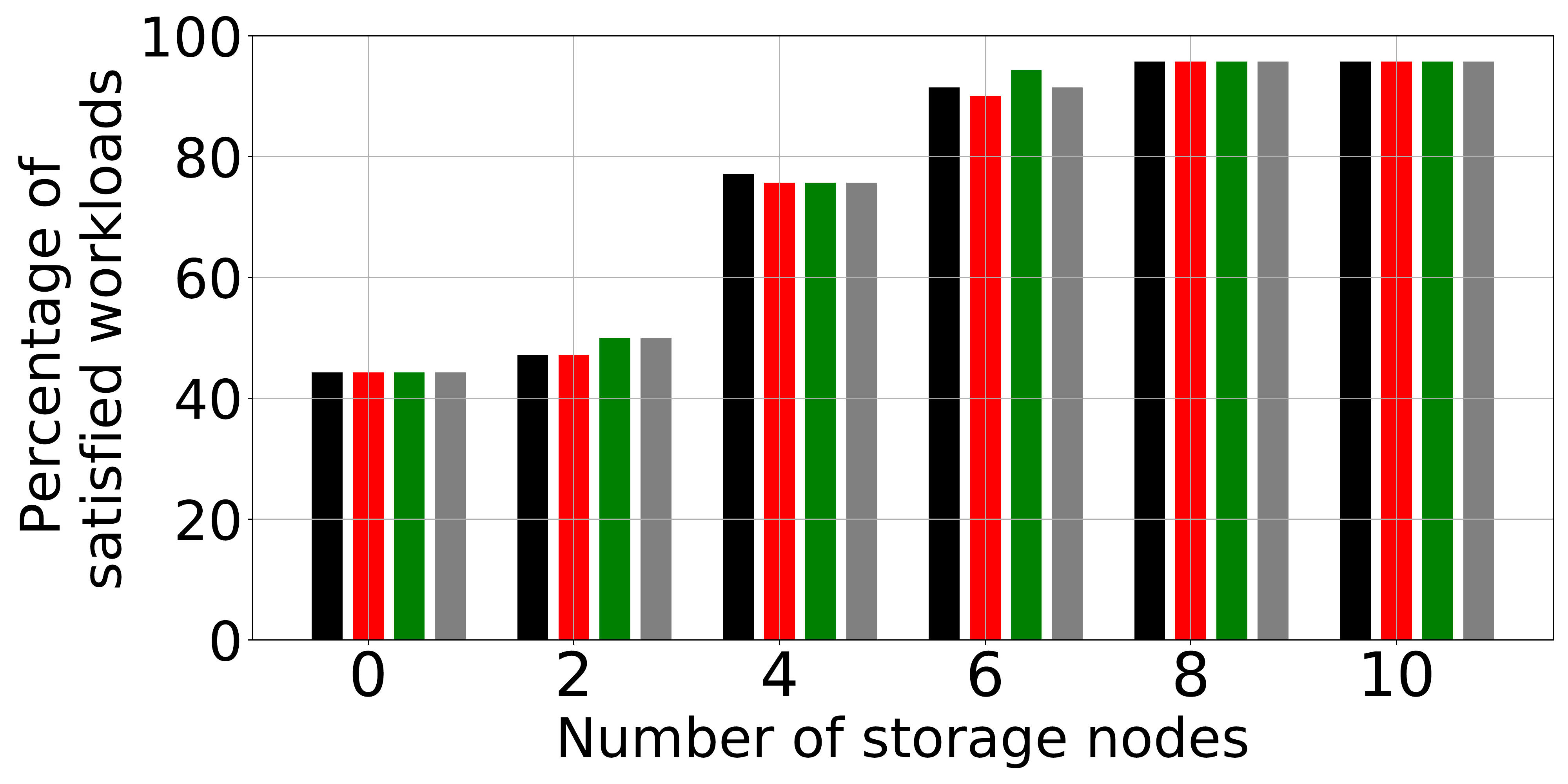}  
    \vspace{-0.06in}
    \caption{$PA(50,2)$}
    \label{fig:satified_in_random3}
\end{subfigure}
\begin{subfigure}{.24\textwidth}
    \centering
    \includegraphics[width=.9\linewidth]{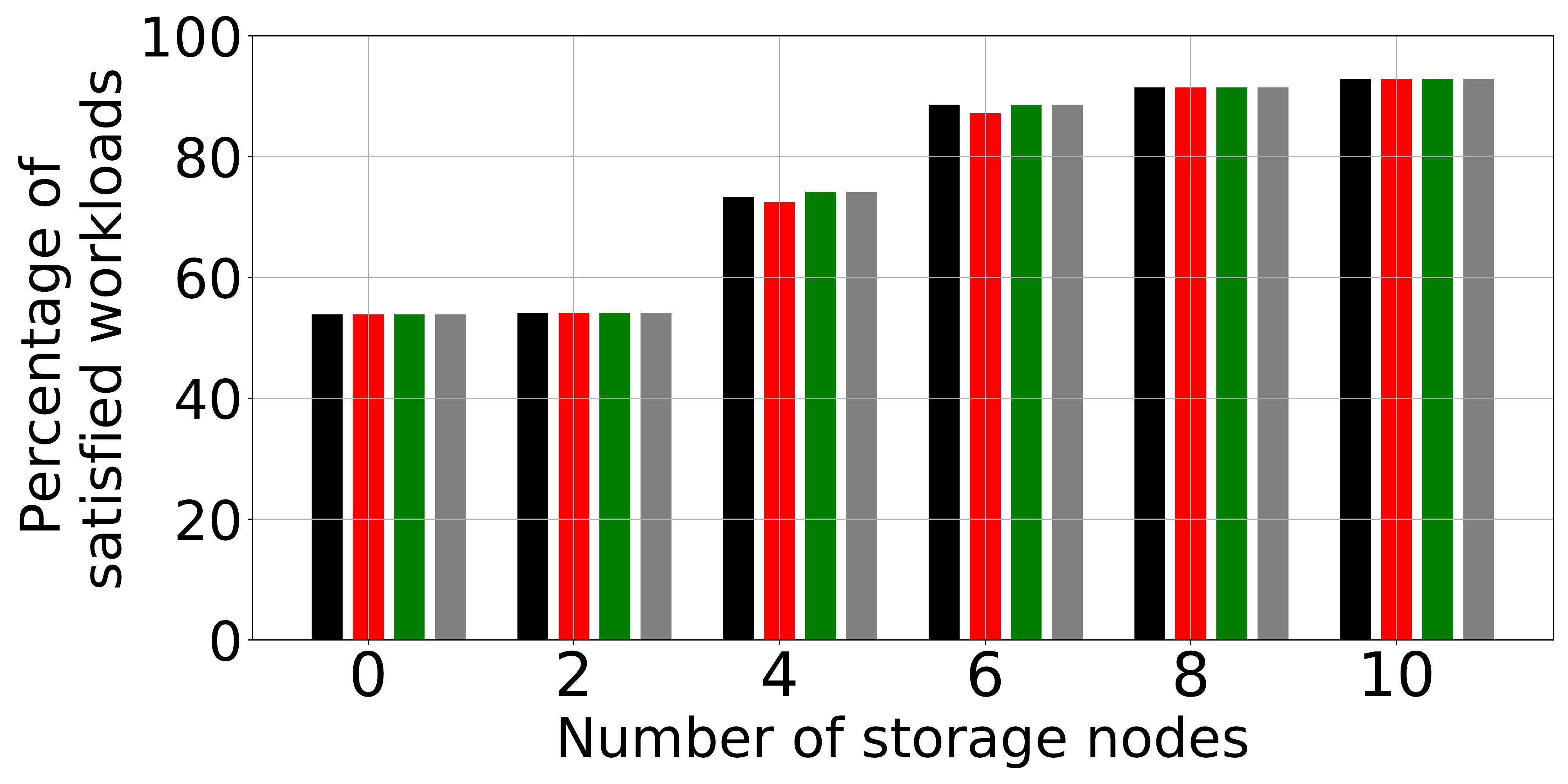} 
    \vspace{-0.06in}
    \caption{$PA(50,3)$}
    \label{fig:satified_in_surfnet}
\end{subfigure}

 \vspace{-0.06in}
\caption{Percentage of satisfied workloads in real (\ref{fig:satified_in_att},\ref{fig:satified_in_abilene}, and \ref{fig:satified_in_surfnet}) and random (\ref{fig:satified_in_random1},\ref{fig:satified_in_random2}, and \ref{fig:satified_in_random3}) topologies with different number of storage nodes. $B_s$ for all servers is $12,000$, $\Delta$ is 20 seconds, and the fidelity threshold for requests is 0.8.   \label{fig:affect_of_storage_and_life_time_of_pick_handling}}
\end{figure*}

In this section, we experimentally evaluate QONs to answer the following questions: (1) Do QONs help to handle demand spikes from users and help to serve more EPR pairs in the network? (2) Do QONs significantly drive down entanglement request service delays by storing and moving EPR pairs closer to users? (3) How do different strategies for storage node selection in the network affect the answers to two previous questions?

We conduct the experiments using the IBM CPLEX solver \cite{cplex}. 

\begin{table}
    \centering
    \setlength{\tabcolsep}{4.9pt}
    \small
    \begin{tabularx}{\columnwidth}{l l l l l l}
        \toprule
        Param & $\Delta$ & $|T|$ & $c(u,v)$ & Link Fidelity& $B_s$\\ 
        \midrule
        Value & $20$s&$10$&Unif$[200,1400]$&Unif$[0.96,0.99]$&$12000$\\
        \bottomrule
    \end{tabularx}
    \vspace{-0.08in}
    \caption{\label{tab:parameters_value} Parameters used in our experiments.}
\end{table}

\myparab{Network topologies:} In our evaluation, we use the topologies of four real networks including the Dutch SURFnet network, taken from the Internet topology zoo \cite{knight2011internet}, ATT and IBM network topologies taken  from \cite{Teavar_source_code}, Abilene \cite{Abilene}, and different randomly generated topologies based on preferential attachment model (power law graph) and \textit{Erdos Renyi} graph \cite{barabasi1999emergence}. We use $G(n,p)$ to refer to the \textit{Erdos Renyi} graphs where $n$ is the number of nodes and $p$ is the probability for edge creation. We use $PA(n,m)$ to refer to the preferential attachment model where $n$ is the number of nodes and $m$ is the number of edges to attach from a new node to existing nodes. We use python \textit{networkx} library \cite{SciPyProceedings_11} to generate random topologies.

Table \ref{tab:parameters_value} shows the value of different parameters used in our experiments. Table \ref{table:topology_charecteristics} specifies the characteristics of the topologies. We average the degree and the diameter over 5 randomly generated topologies for $G(n,p)$ and $PA(n,m)$ where $p=0.1,0.05$ and $m=2,3$. In all topologies, unless mentioned, we consider at most one shortest path (based on hop count) between each pair of users, each pair of storage servers, and each user and each storage server.

\myparab{Purification and Swapping:} We assume all noisy mixed entangled states in the network are Werner states \cite{werner1989quantum}. In our experiments we use the recurrence based purification scheme as explained in Section \ref{sec:Preliminaries}. In particular, we use the DEJMPS protocol \cite{Deutsch96} for purification. The values of $p_k$ and $k_{max}$ in \eqref{eq:purification_avg} can be determined from the results presented in \cite{Deutsch96}. When a node performs an (noise-free) entanglement swap operation between two EPR-pairs with fidelities $F_1$ and $F_2$, the fidelity of the resulting state is $\frac{1}{4}+\frac{3}{4}*(\frac{4F_1-1}{3})(\frac{4F_2-1}{3})$ \cite{briegel1998quantum}. We compute the basic fidelity of a path in the same way.

\myparab{Storage selection schemes:}
We select the storage servers in the network based on two schemes. In \textit{Random} scheme, we select storage nodes randomly. In \textit{Degree} scheme, nodes with the higher degrees are selected first as storage servers.

\begin{table}[t]
\footnotesize
\begin{small}
    \centering
    \small
    \footnotesize
    \begin{tabularx}{\columnwidth}{X c c c c }
      \toprule
      {Topologies} &   {$|V|$} & {$|E|$}&Avg. Degree &Diameter \\
      \midrule
      {ATT} &   {25} & {112}&4.4 &{5} \\ 
      \midrule
      {Abilene}  & {12}  & {30} & 2.5&{5} \\
      
      \midrule
      {IBM} & {17}  &  {46} &2.7& {6} 
    \\
    \midrule
      {SURFnet} & {50}  &  {68} &2.7& {11} 
         \\
    \midrule
      {$G(50,0.0.5)$} & {50}  &  {67} &2.8& {9.3} 
    \\
    \midrule
      {$G(50,0.1)$} & {50}  &  {123} &4.9& {5.3} 
 
          \\
    \midrule
      {$PA(50,2)$} & {50}  &  {96} &3.8& {4.9} 
    \\
    \midrule
      {$PA(50,3)$} & {50}  &  {141} &5.6& {4} 

    \\
      \bottomrule
    \end{tabularx}
    \vspace{-0.05in}
    \caption{\label{table:topology_charecteristics} Properties of network topologies used in Simulations.}
    \vspace{-0.1in}
    \end{small}
\end{table}

\eat{
 \begin{figure}[t]
\centering
    \includegraphics[width=6cm]{img/demands.pdf}
  \caption{One example of demand traces used in experiments \label{fig:workload_traces}}
\end{figure}}

\myparab{Workload:} We consider $6$ user pairs for each network topology and generate the demands to create entanglements between the user pairs for each time interval using the spike model of \textit{tgem} library \cite{tmgen}. The spike model takes the number of user pairs, number of time intervals, number of user pairs that would have a spike in their demands, and a mean value for the spike as inputs and generates the demands between each pair of users for each time interval. In our experiments, at each time interval, three user pairs can have a spike in their demands. A workload includes demands from all user pairs and their fidelity requirements over $T$ time intervals. Since each topology has a different capacity for serving entanglements, we normalize the generated workloads of each topology based on the capacity of that topology. We first compute the capacity of each topology as the maximum EPR rate that it can generate with a fidelity threshold $0.75$ to different sets of $6$ user pairs. If the capacity of topology, with this definition, is $c$, we set the mean value of each spike to $c/3$ as we have three spikes in each time interval in our workload generation module.

\subsection{Handling demand spikes}
\label{exp:work_load_satisfying}
We first evaluate the robustness of QONs in handling unexpected spikes in user demands. We say a workload is satisfied if the network can serve the demands of all users in all time intervals while fulfilling their fidelity requirements. For each network, we measure the percentage of $200$ generated workloads that the network can satisfy and plot them as a function of number of storage nodes in the network. 

Figure \ref{fig:affect_of_storage_and_life_time_of_pick_handling} shows the percentage of satisfied workloads among all workloads for real and randomly generated topologies. In each sub-figure, the first keyword in the legend indicates the value of $h$ that represents the EPR pair storage lifetime and the second keyword represents the scheme for storage node selection. The threshold for fidelity requirement of all requests at all time intervals in this experiment is set to 0.8 and the values of $\Delta$ and $B_s$ (for each storage server $s$) are 20 seconds and 12,000 EPR pairs respectively.

A couple observations are noteworthy. As expected, the number of satisfied workloads increases with the number of storage nodes in the network and the highest value observed is around $92\%$ with $10$ storage nodes in most topologies. There is almost no difference between infinite time-intervals for EPR pair lifetime ($h\geq|T|$) and one-time-interval lifetime extreme ($h=1$) except for the case when the number of storage nodes is low in the network. In addition, the percentage of satisfied workloads varies across different topologies. Topologies with a higher number of edges can handle more demand spikes as the paths would use  disjoint links with a higher probability in these networks. Two storage selection schemes produce different results for different topologies. However, in most cases, choosing storage nodes using \textit{Degree} scheme (red and black bars) outperforms the random scheme.

\subsection{Storage utilization}
% \vspace{-0.03in}
We now evaluate the utilization of storage nodes in the network under different target fidelity requirements for end-user requests. For each storage server $s$, if the highest number of EPR pairs that is stored at it among all time intervals of a workload is $v$ and the capacity of the storage is $B_s$, we define the utilization of that storage to be $\frac{v}{B_s}*100$. For each workload, we measure the utilization for all storage servers in the network and compute the average. For each generated workload, we change the fidelity requirement of the requests for $6$ user pairs in the network and measure the storage server utilization.  We use the previous experiment setup for this experiment as well. We conduct this experiment with ATT and $PA(50,2)$ routing topologies.

\begin{figure}

  \begin{subfigure}{.22\textwidth}
    \includegraphics[width=4.0cm]{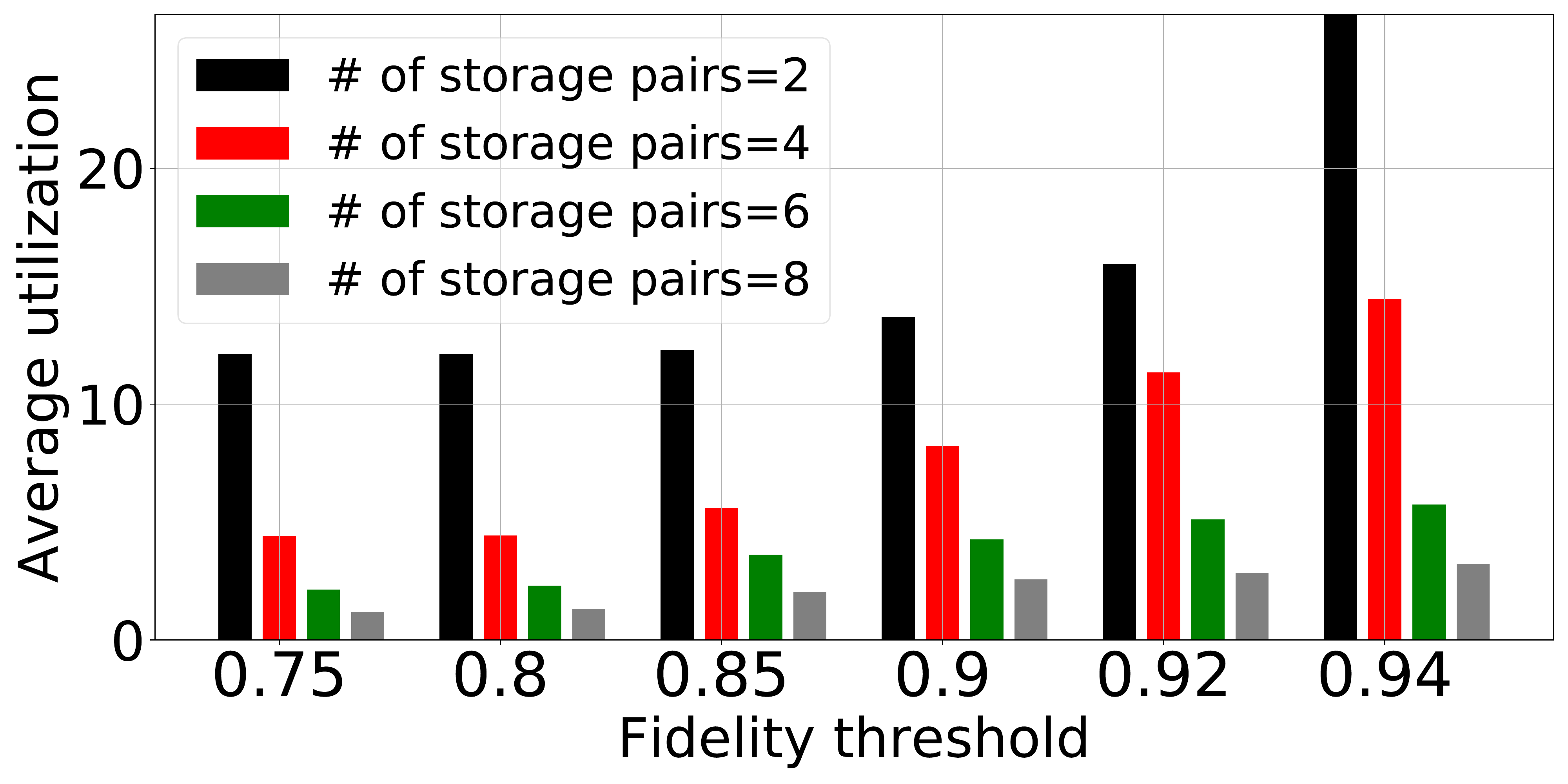}
    \vspace{-0.2in}
  \caption{ATT \label{fig:real_topologies}}
  \end{subfigure}
  \begin{subfigure}{.22\textwidth}
    \includegraphics[width=4.0cm]{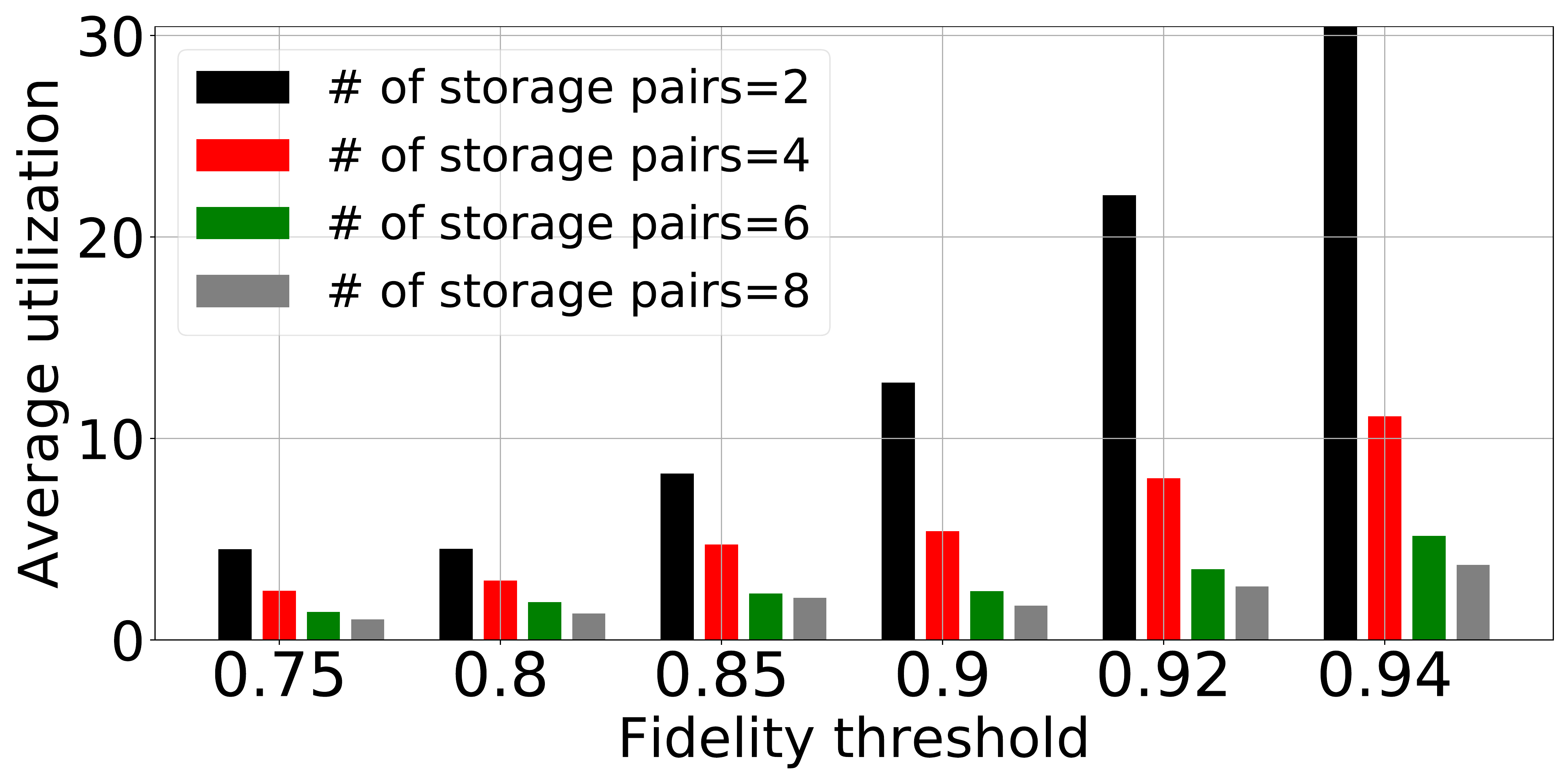}
    \vspace{-0.2in}
    \caption{$PA(50,2)$ \label{fig:random_topologies}}
  \end{subfigure}
\vspace{-0.06in}
  \caption{Storage server utilization under different fidelity requirements for requests for \textit{Degree} storage selection scheme, $B_s=12,000$ and $h \geq |T|$ case.  \label{fig:storage_utilization_exp}}
\end{figure}

Figure \ref{fig:storage_utilization_exp} shows the average storage utilization across different number of storage servers in the network. The storage servers are chosen using \textit{Degree} scheme. As might be expected utilization decreases as the number of servers increases as the load gets spread over more servers. By increasing the fidelity requirement of requests, the utilization is increased as we need more EPR pairs for the purification scheme ans more EPR pairs are stored at storages for this purpose.   
% and across all time intervals and workloads

\begin{figure}

  \begin{subfigure}{.22\textwidth}
    \includegraphics[width=3.7cm]{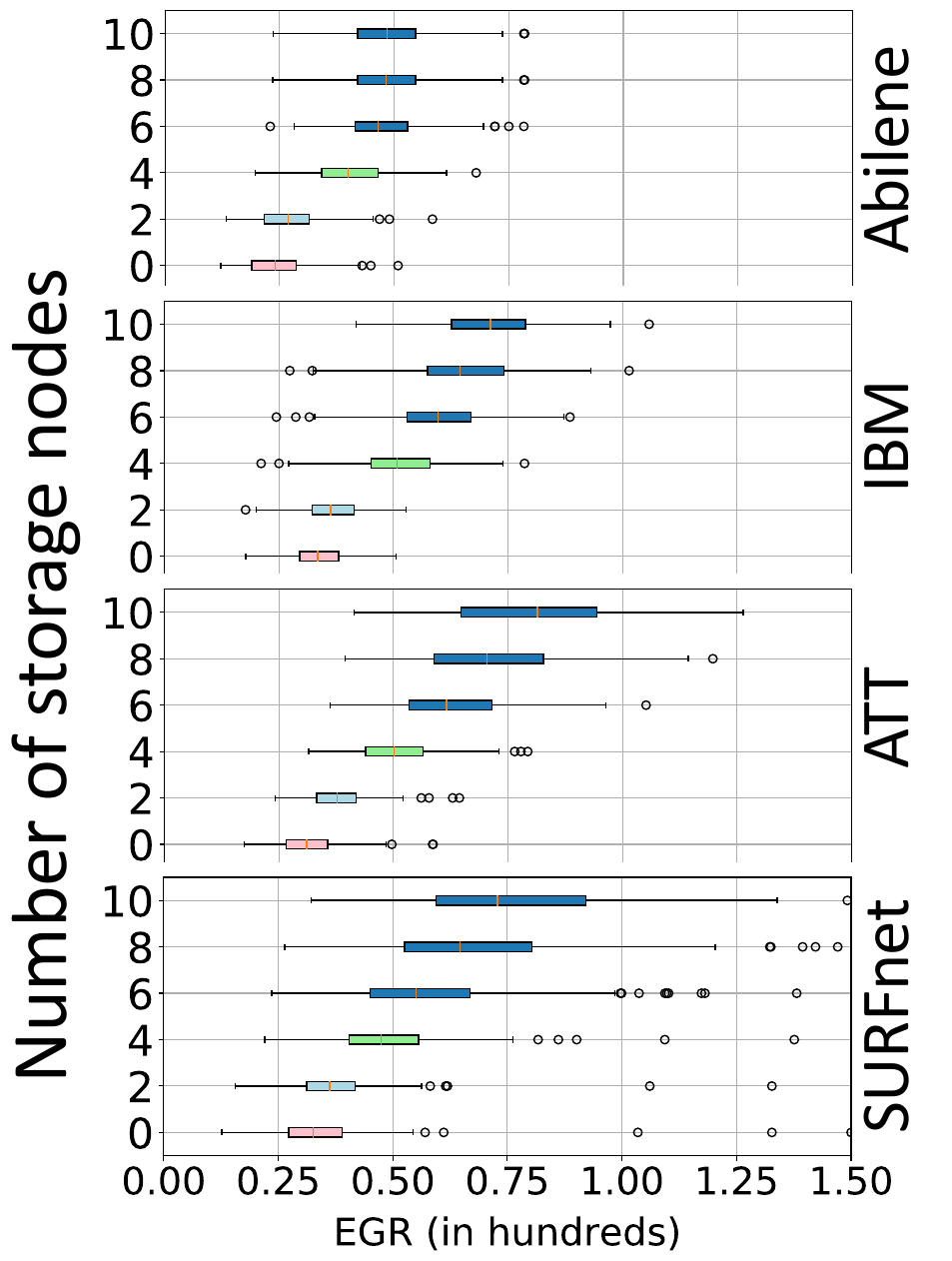}
    \vspace{-0.06in}
  \caption{Real topologies \label{fig:real_topologies}}
  \end{subfigure}
  \begin{subfigure}{.22\textwidth}
    \includegraphics[width=3.7cm]{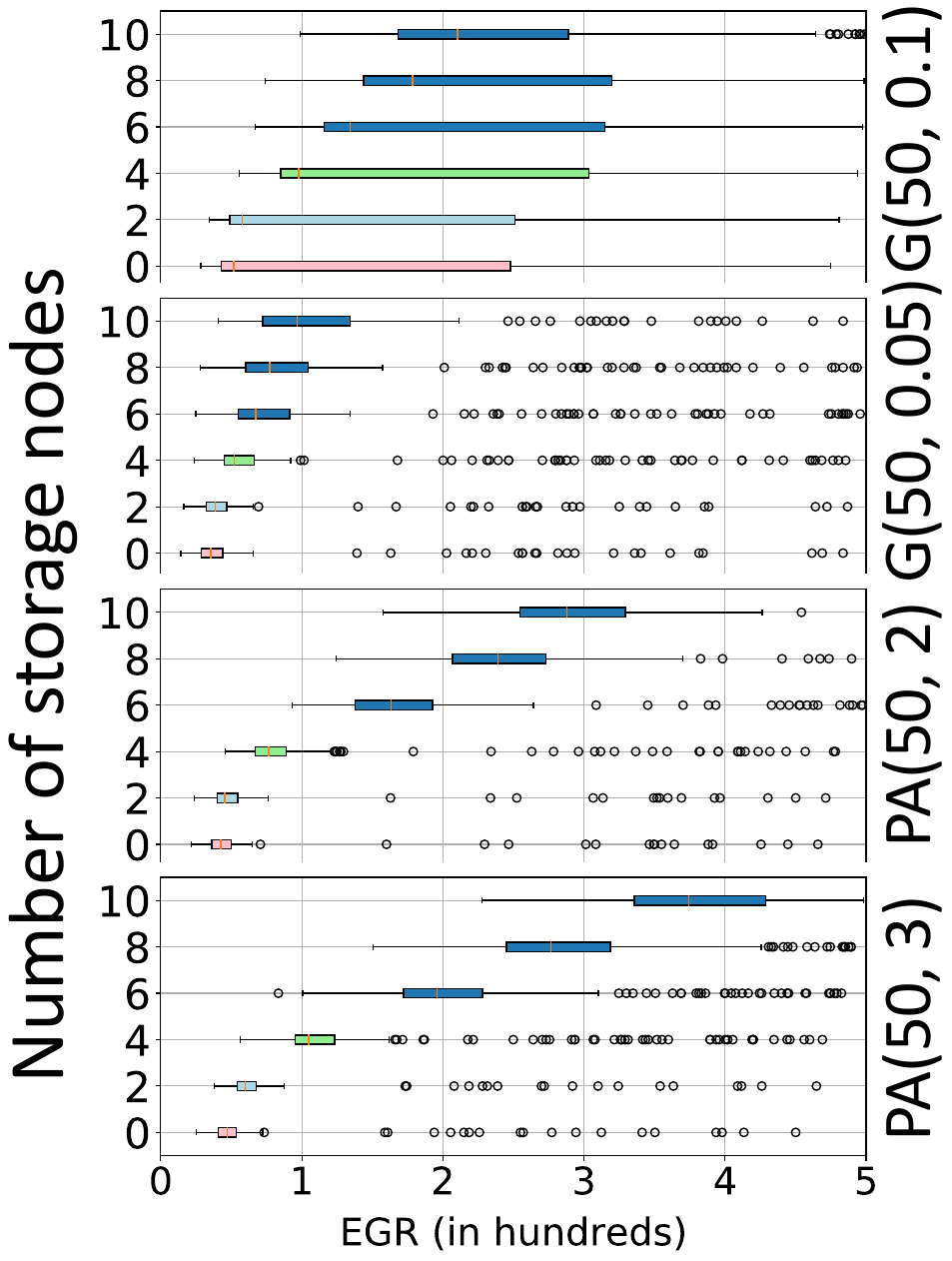}
    \vspace{-0.06in}
    \caption{Random topologies \label{fig:random_topologies}}
  \end{subfigure}
%   \vspace{-0.09in}
  \caption{Maximizing EGR in different topologies. Storage selection scheme is \textit{Degree} and $h\geq |T|$.  The fidelity threshold is 0.8. \label{fig:EGR_over_dynamic_weights}}
\end{figure}

 \begin{figure}[t]
\centering
    \includegraphics[width=5cm]{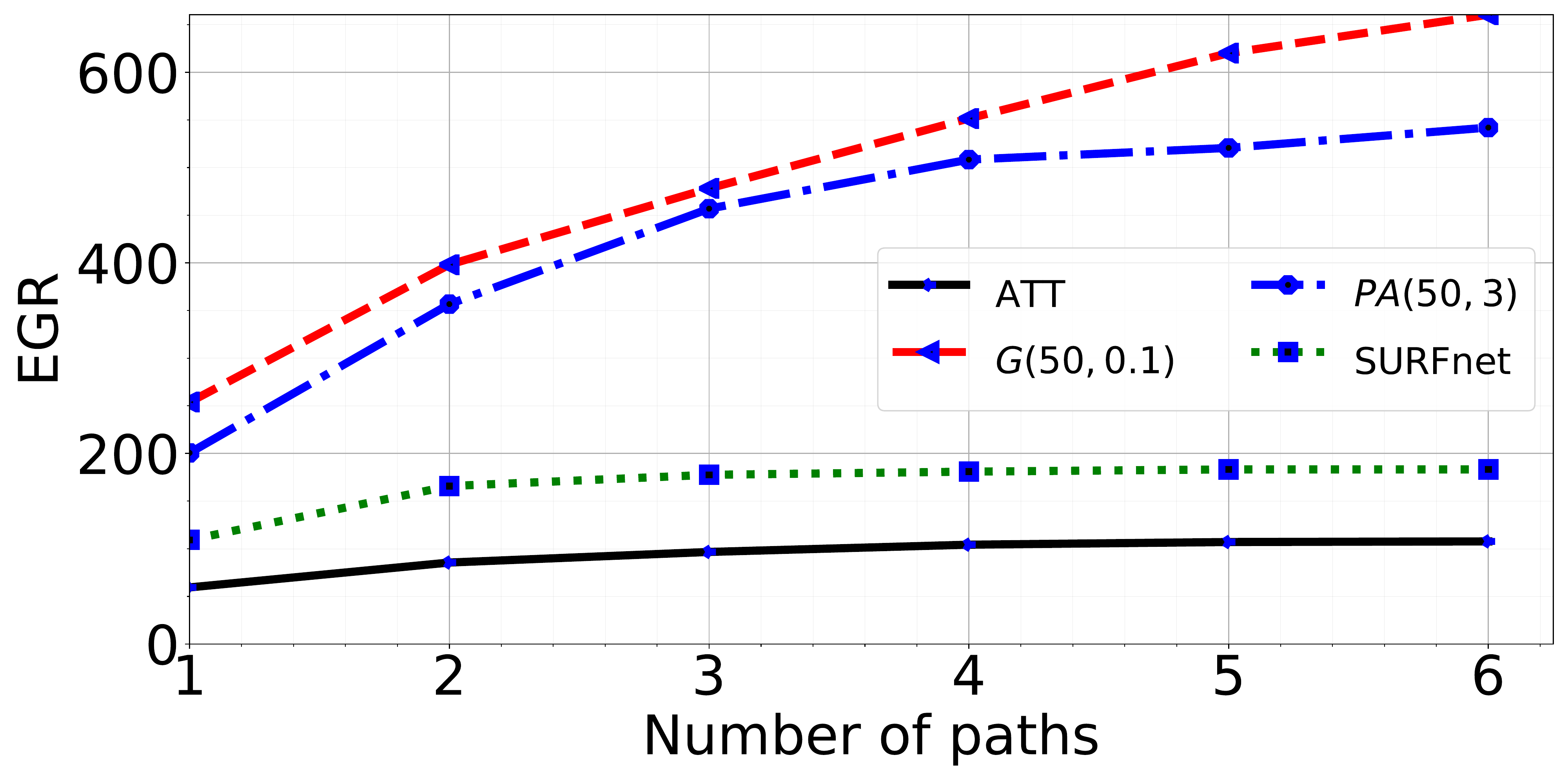}
    \vspace{-0.06in}
  \caption{EGR as a function of number of paths. \label{fig:path_diversity}}
\end{figure}

\begin{figure*}[t]
\centering
\begin{subfigure}{.24\textwidth}
    \centering
    \includegraphics[width=.95\linewidth]{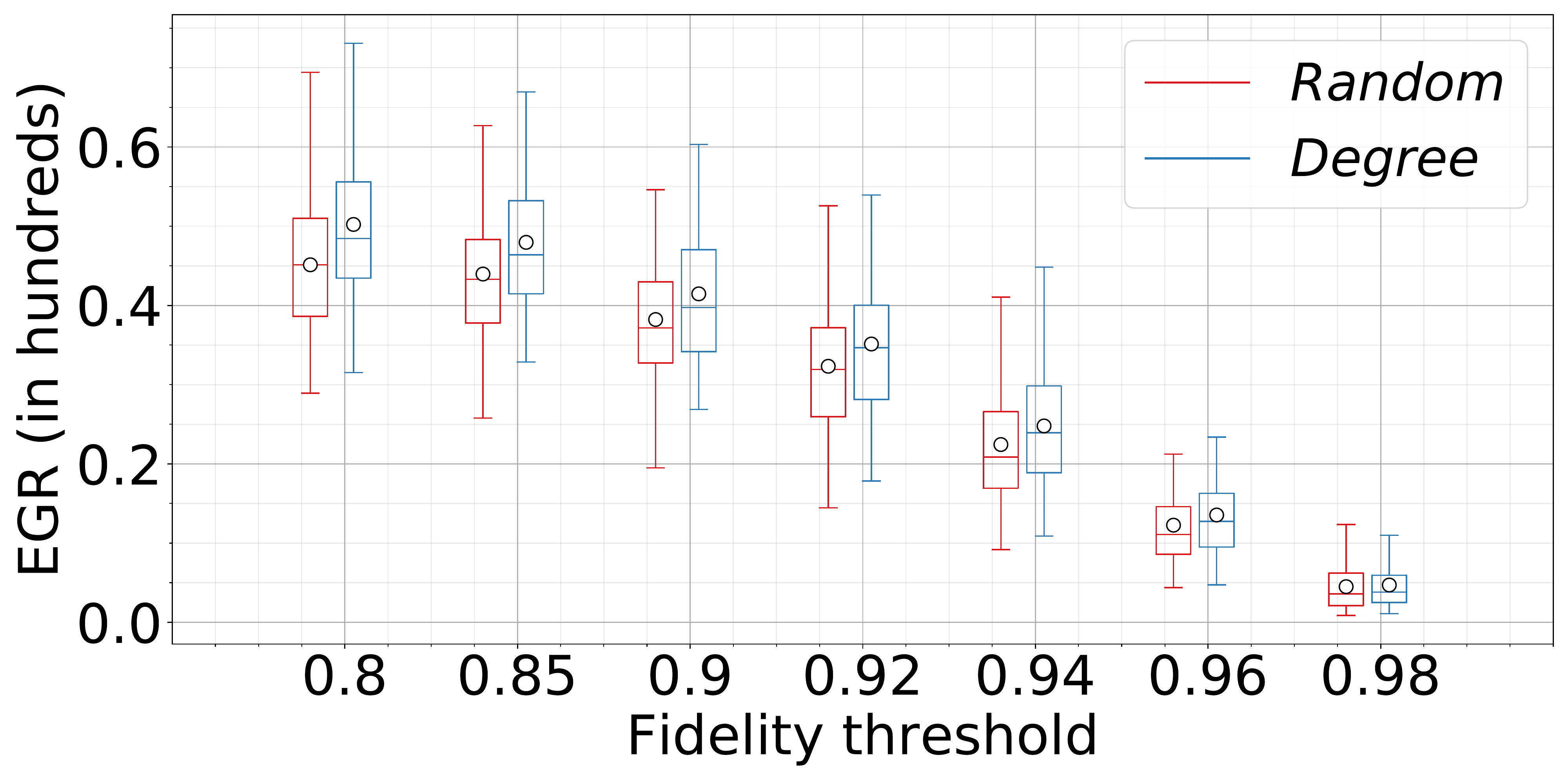}
    \vspace{-0.06in}
    \caption{ATT}
    \label{fig:satified_in_att2}
\end{subfigure}
\begin{subfigure}{.24\textwidth}
    \centering
    \includegraphics[width=.95\linewidth]{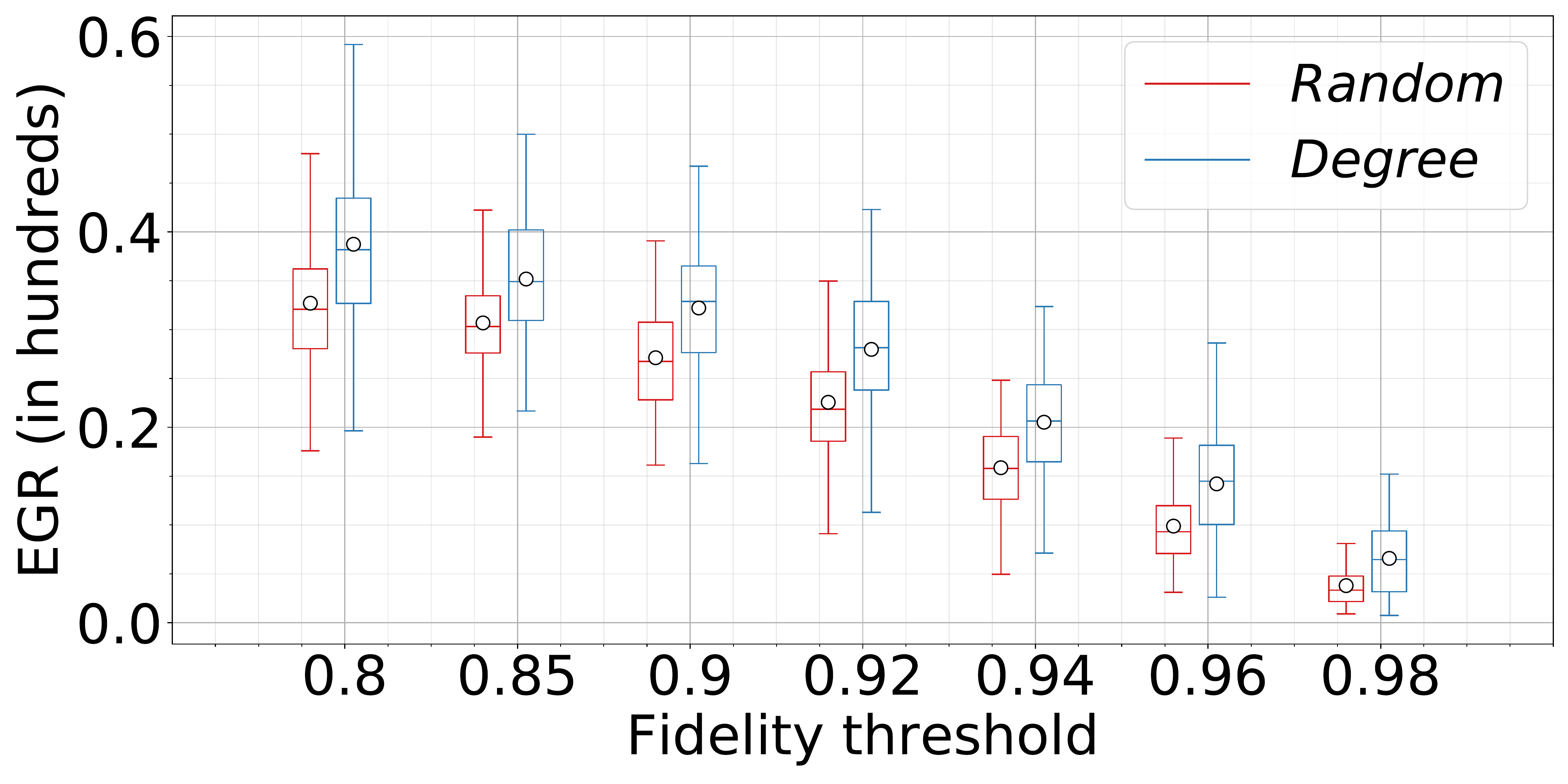}  
    \vspace{-0.06in}
    \caption{Abilene}
    \label{fig:satified_in_abilene2}
\end{subfigure}
\begin{subfigure}{.24\textwidth}
    \centering
    \includegraphics[width=.95\linewidth]{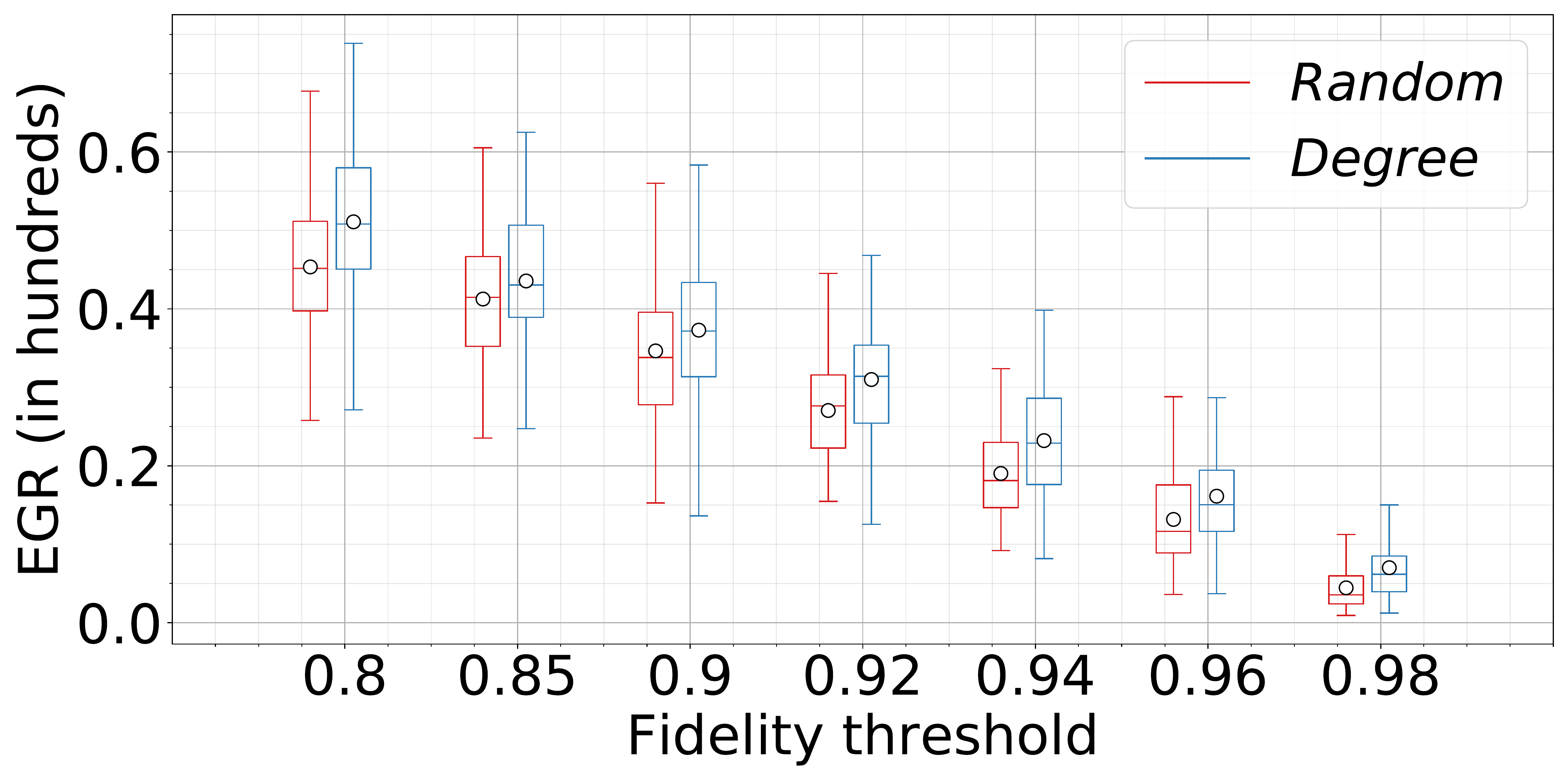} 
    \vspace{-0.06in}
    \caption{IBM}
    \label{fig:satified_in_surfnet2}
\end{subfigure}
\begin{subfigure}{.24\textwidth}
    \centering
    \includegraphics[width=.95\linewidth]{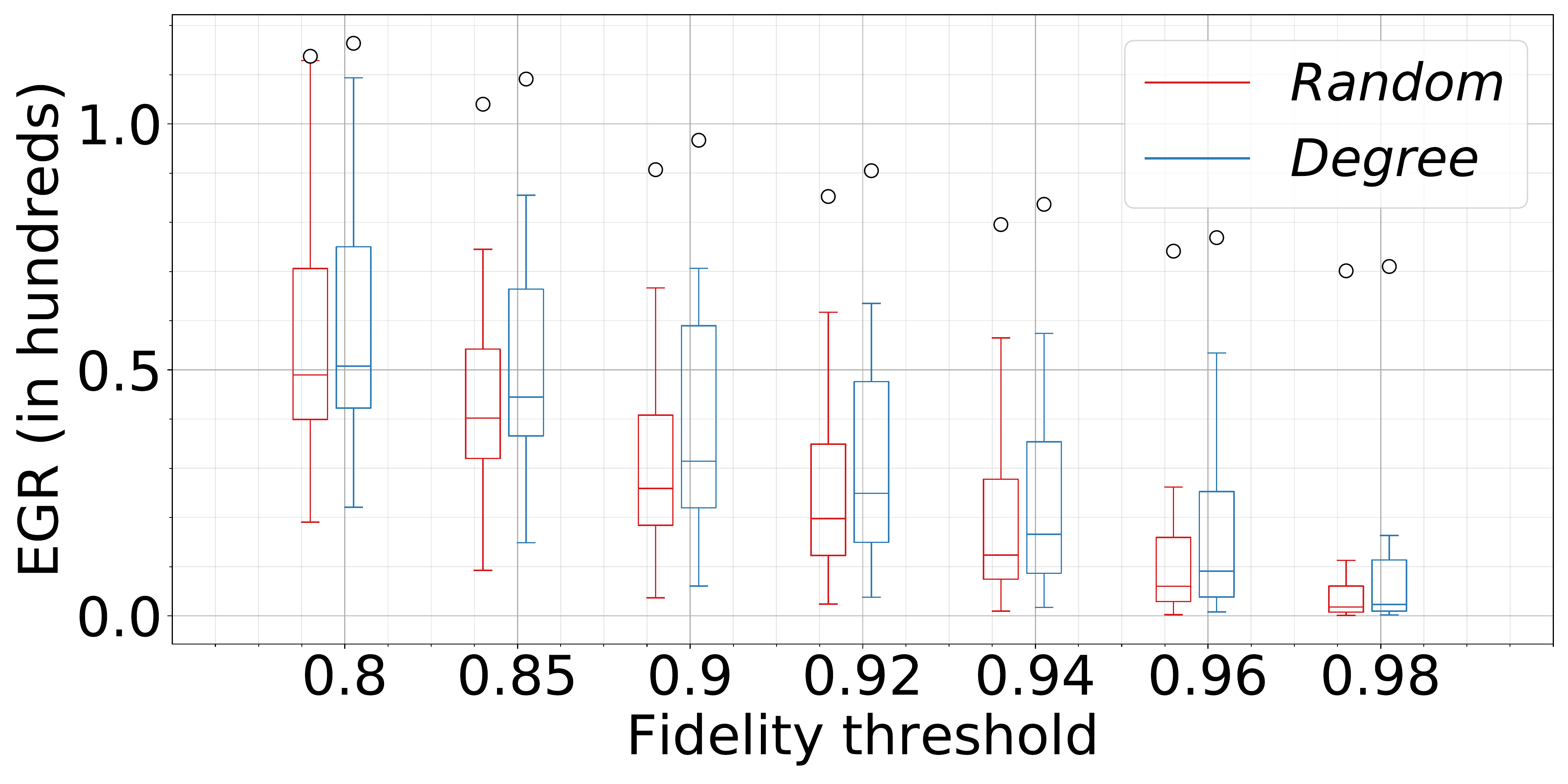}
    \vspace{-0.06in}
    \caption{SURFnet}
    \label{fig:satified_in_surfnet2}
\end{subfigure}

\begin{subfigure}{.24\textwidth}
    \centering
    \includegraphics[width=.95\linewidth]{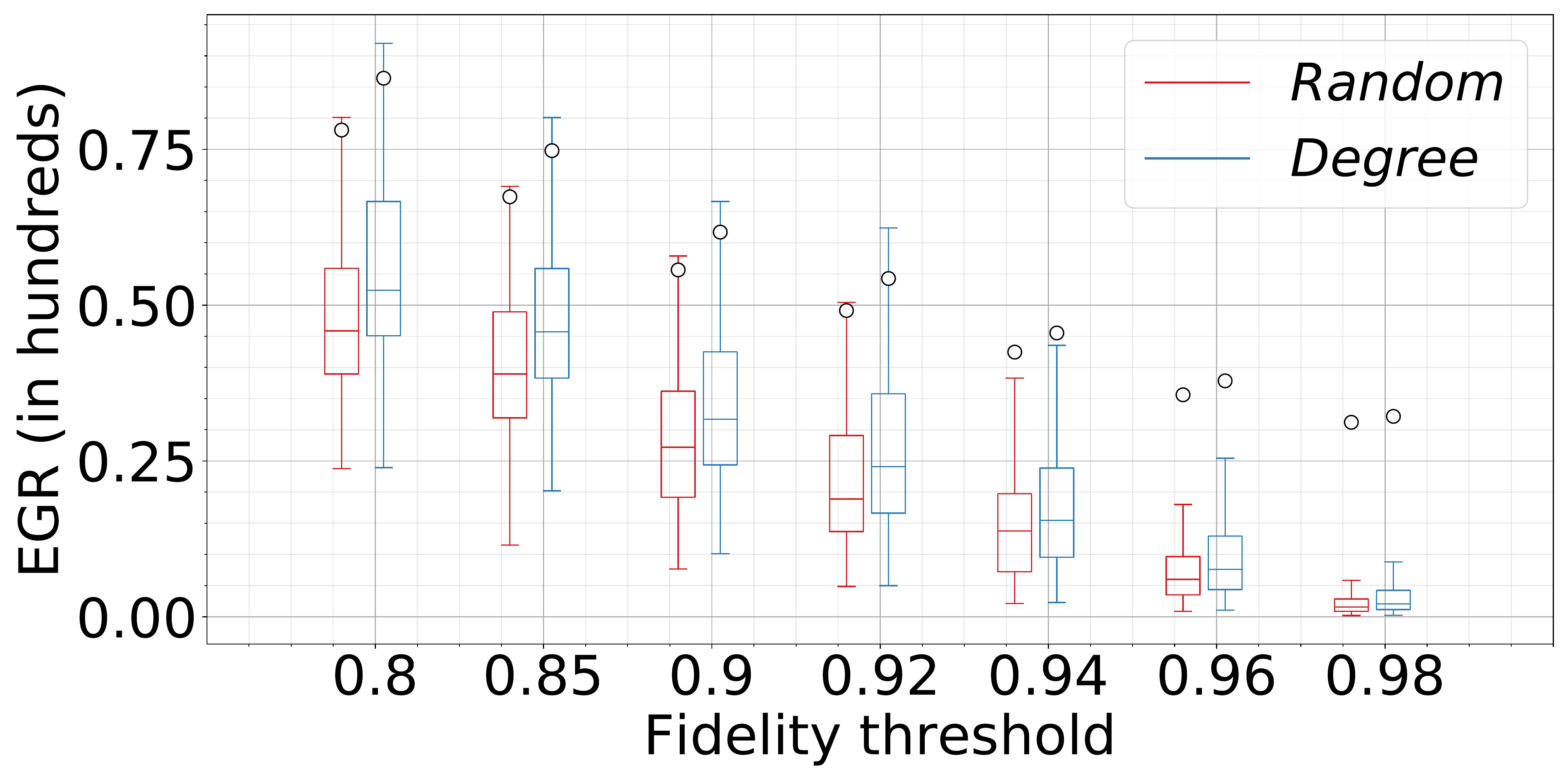}  
    \vspace{-0.06in}
    \caption{$G(50,0.05)$}
    \label{fig:satified_in_random22}
\end{subfigure}
\begin{subfigure}{.24\textwidth}
    \centering
    \includegraphics[width=.95\linewidth]{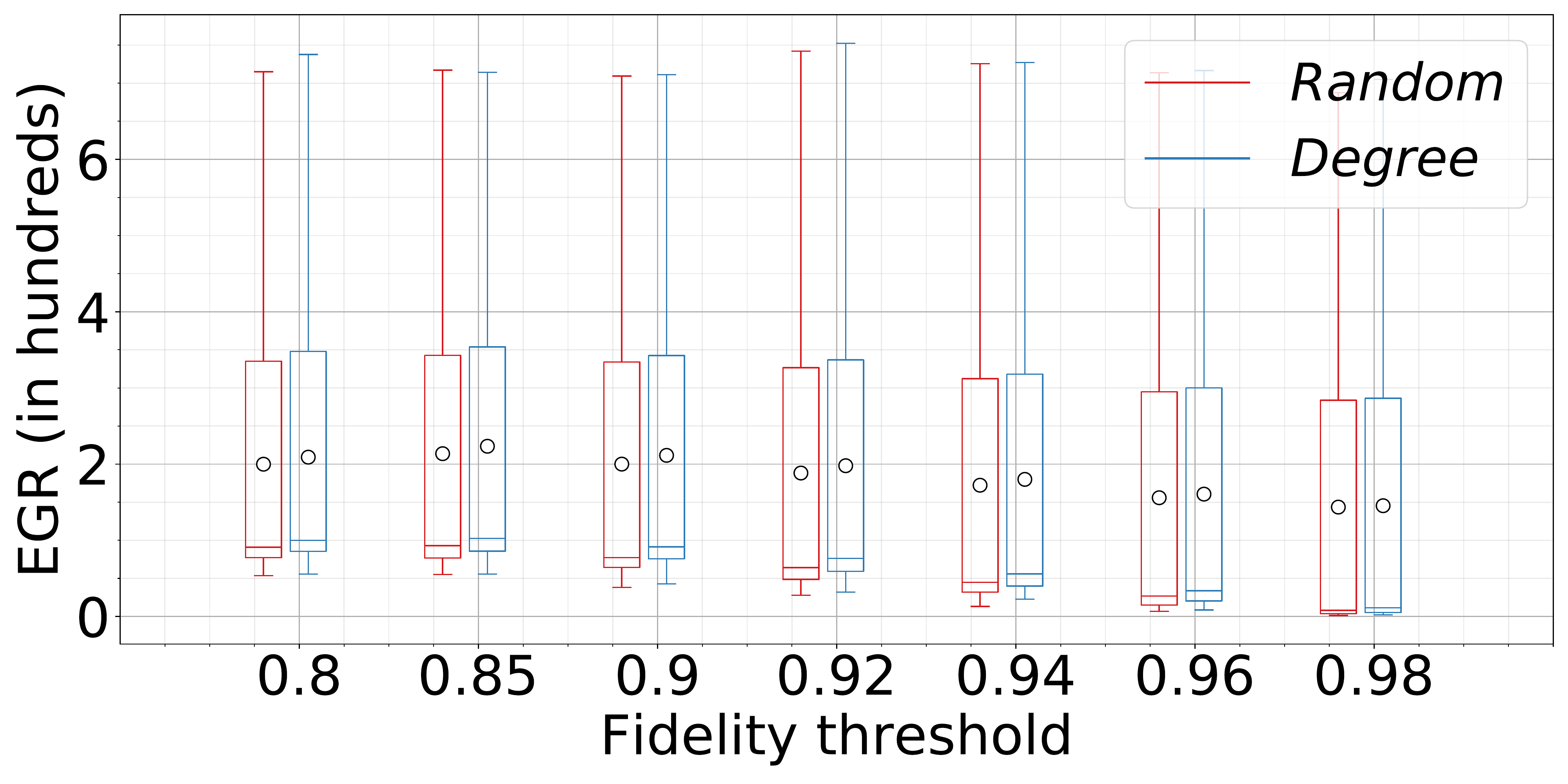}  \vspace{-0.06in}
    \caption{$G(50,0.1)$}
    \label{fig:satified_in_random12}
\end{subfigure}
\begin{subfigure}{.24\textwidth}
    \centering
    \includegraphics[width=.95\linewidth]{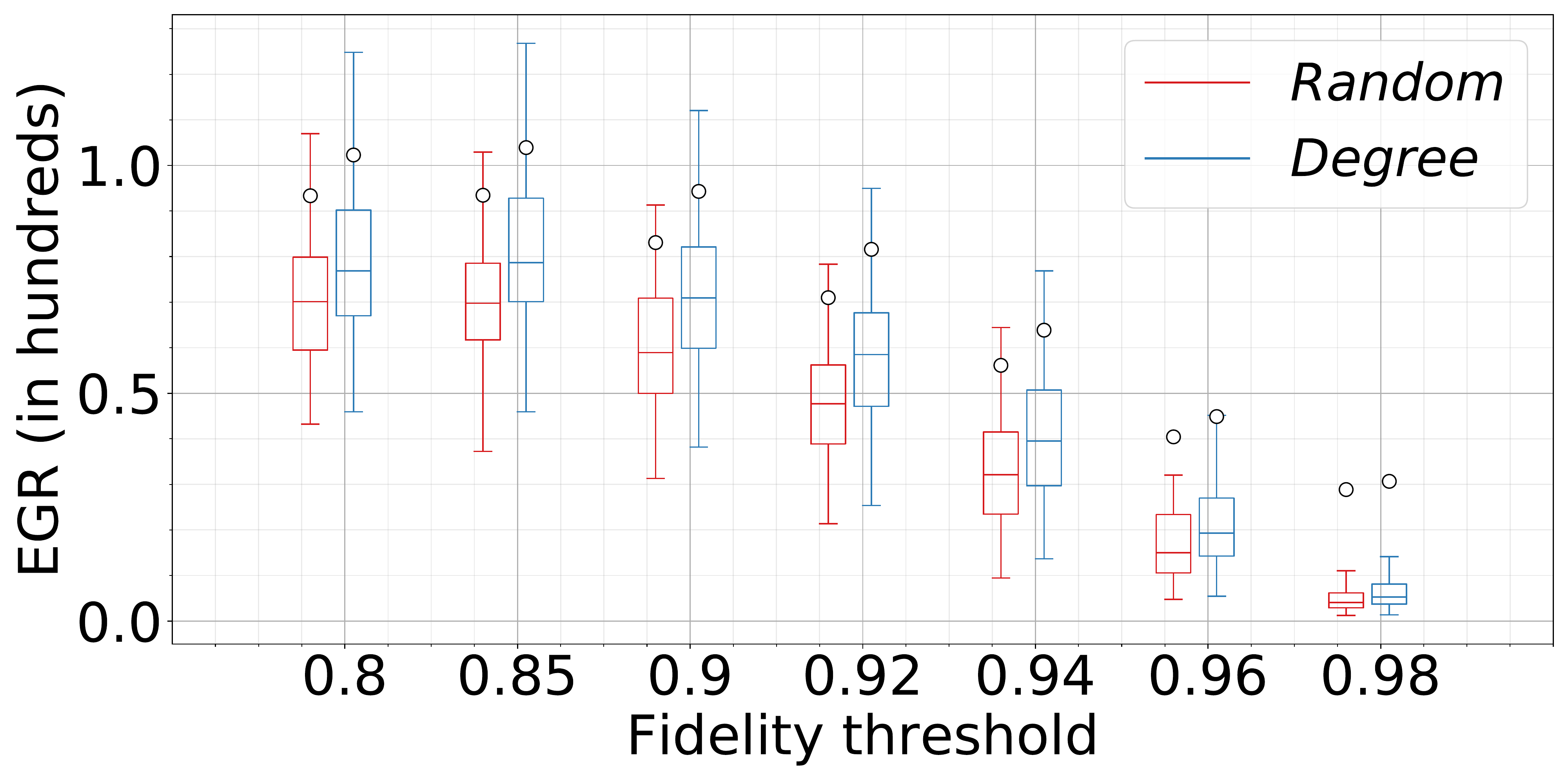}
    \vspace{-0.06in}
    \caption{$PA(50,2)$}
    \label{fig:satified_in_random32}
\end{subfigure}
\begin{subfigure}{.24\textwidth}
    \centering
    \includegraphics[width=.95\linewidth]{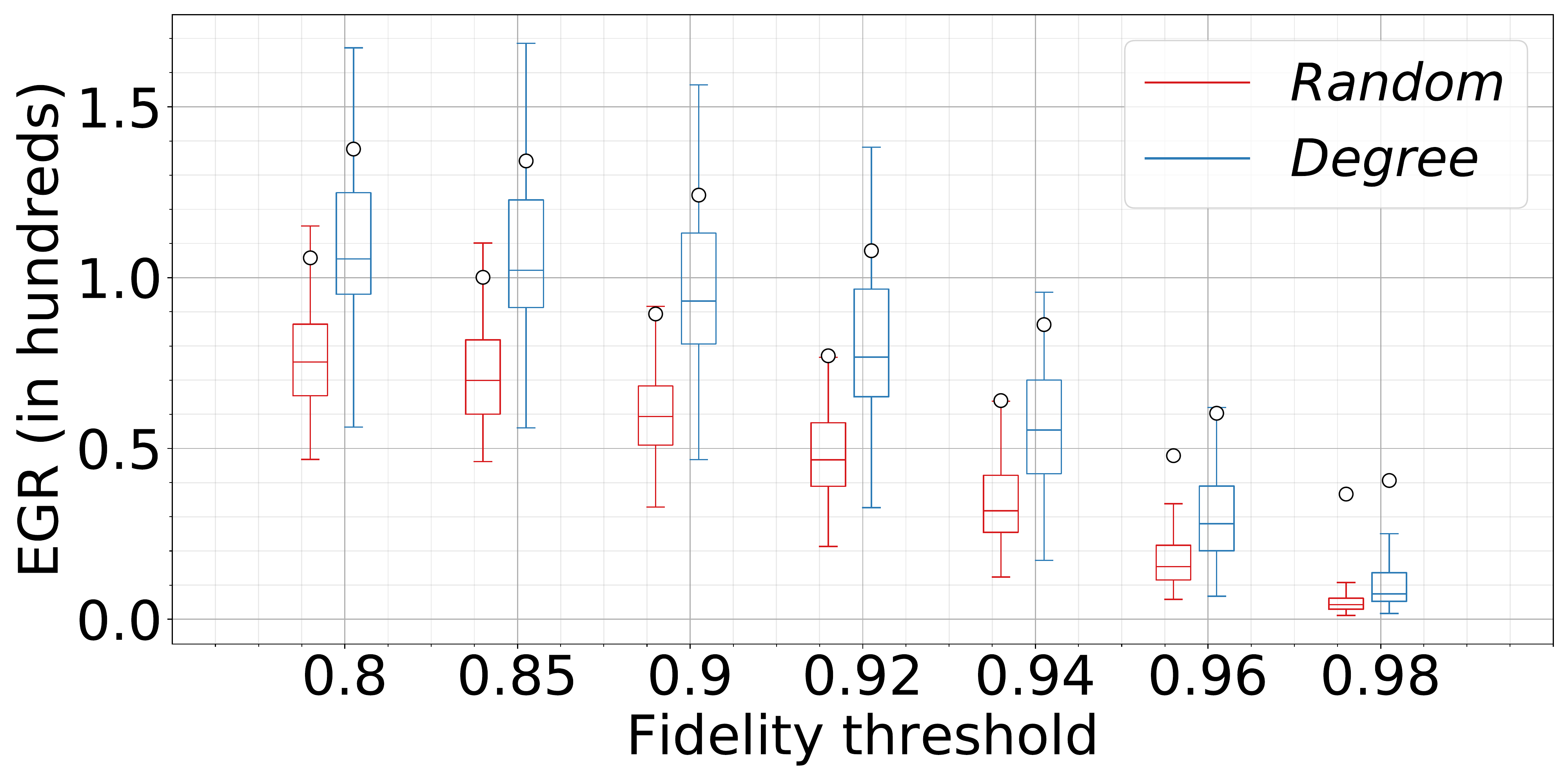}  
    \vspace{-0.06in}
    \caption{$PA(50,3)$}
    \label{fig:satified_in_random222}
\end{subfigure}

\vspace{-0.06in}
\caption{EGR in real (top) and random (below) topologies as a function of fidelity threshold using 4 storages nodes selected using \textit{Random} and \textit{Degree} approaches. For this figure, we have $h\geq |T|$ and $\Delta=20$ seconds and $B_s=12000$.\label{fig:egr_as_a_function_of_fidelity_threshold}}
\end{figure*}

\subsection{Maximizing Weighted Entanglement Generation Rate}
\label{exp:EGR_maximizing}

In this experiment, we evaluate the maximum rate that a QON can serve to a set of users. We assume the set of users for all time intervals is fixed but the weight for each user is changing. We set the weight of each user pair at each time interval from the range $[0,1]$. We assume the fidelity threshold for all delivered EPR pairs is equal and is $0.8$. We conduct this experiment with the \textit{Degree} scheme for storage node selection in the network.

Figure \ref{fig:EGR_over_dynamic_weights} shows a box plot of EGR as a function of the number of storage nodes in the network with $6$ user pairs over $400$ generated workloads of $10$ consecutive time intervals for different topologies. Unsurprisingly, EGR is an increasing function of the number of storage nodes. However, the results for EGR in each topology depends on the average degree and number of edges in the network that can effect the paths used by each user pair. With more edges in the network, it is more likely that different paths would use different links  in comparison to the case that the network has a smaller degree and number of edges. When multiple user pairs use a set of paths that share a link, all user pairs are limited to the capacity of that link. In our evaluation, we found that in $50$ percentage of workloads, in real topologies, an edge is being used by an average of $2.1$ user pairs. However, this value is $1.2$ for random topologies. 

In Figure \ref{fig:path_diversity}, we check the effect of increasing the number of paths between each user pair and storage servers on the EGR of the network. The $y$-axis is the average EGR over $200$ workloads with each workload including $6$ user pairs each having a different weight value for $10$ time intervals. We have plotted the results for only the $h\geq|T|$ case. In this experiment, we select four storage nodes using \textit{Degree} scheme and the fidelity threshold for all served EPR pairs is $0.8.$ Topologies with larger degrees and number of nodes benefit from increasing the number of paths. As explained in the previous paragraph, having more edges also provide more disjoint links between selected paths. 

\subsection{EGR as a function of fidelity threshold}

In another experiment, we measure EGR in different topologies as a function of the fidelity of the delivered EPR pairs. By increasing the fidelity threshold for delivered EPR pairs, we expect more resources will be consumed for purification and that the EGR will decrease. We assume there are four storage servers selected using \textit{Random} and \textit{Degree} schemes in each network and there are $6$ user pairs in the network. The capacity of each storage server is $12,000$ EPR pairs and the edge capacity and fidelity is chosen as the first experiment (exp \ref{exp:work_load_satisfying}). In this experiment, we assume $h\geq |T|$.

Figure \ref{fig:egr_as_a_function_of_fidelity_threshold} shows EGR as a function of fidelity threshold for different topologies. The higher the fidelity threshold, the more resources used for purification and hence lower the EGR. We find that using \textit{Degree} scheme for storage node selection outperforms the \textit{Random} scheme. One reason is that when nodes with higher degrees are selected as storage servers, more users can connect to them via disjoint paths and when not disjoint, there are fewer shared links. We omit the results for the case $h = 1$ as there is little difference from the results for the case $h\geq|T|$. The reason is that the infinite lifetime extreme would outperform the one-time interval extreme only when the weight of users in the next two or more time intervals is larger than the weight of users in the current time interval. Since we set the weight of users at each time interval randomly, this is unlikely to happen.

\subsection{Reducing request service delay}
In this section, we evaluate how service delays are reduced using a QON. We measure the delay of an end-to-end request in the network as the number of entanglement swaps required to establish the end-to-end entanglement.

\begin{figure*}[t]
\centering
\begin{subfigure}{.24\textwidth}
    \centering
    \includegraphics[width=.95\linewidth]{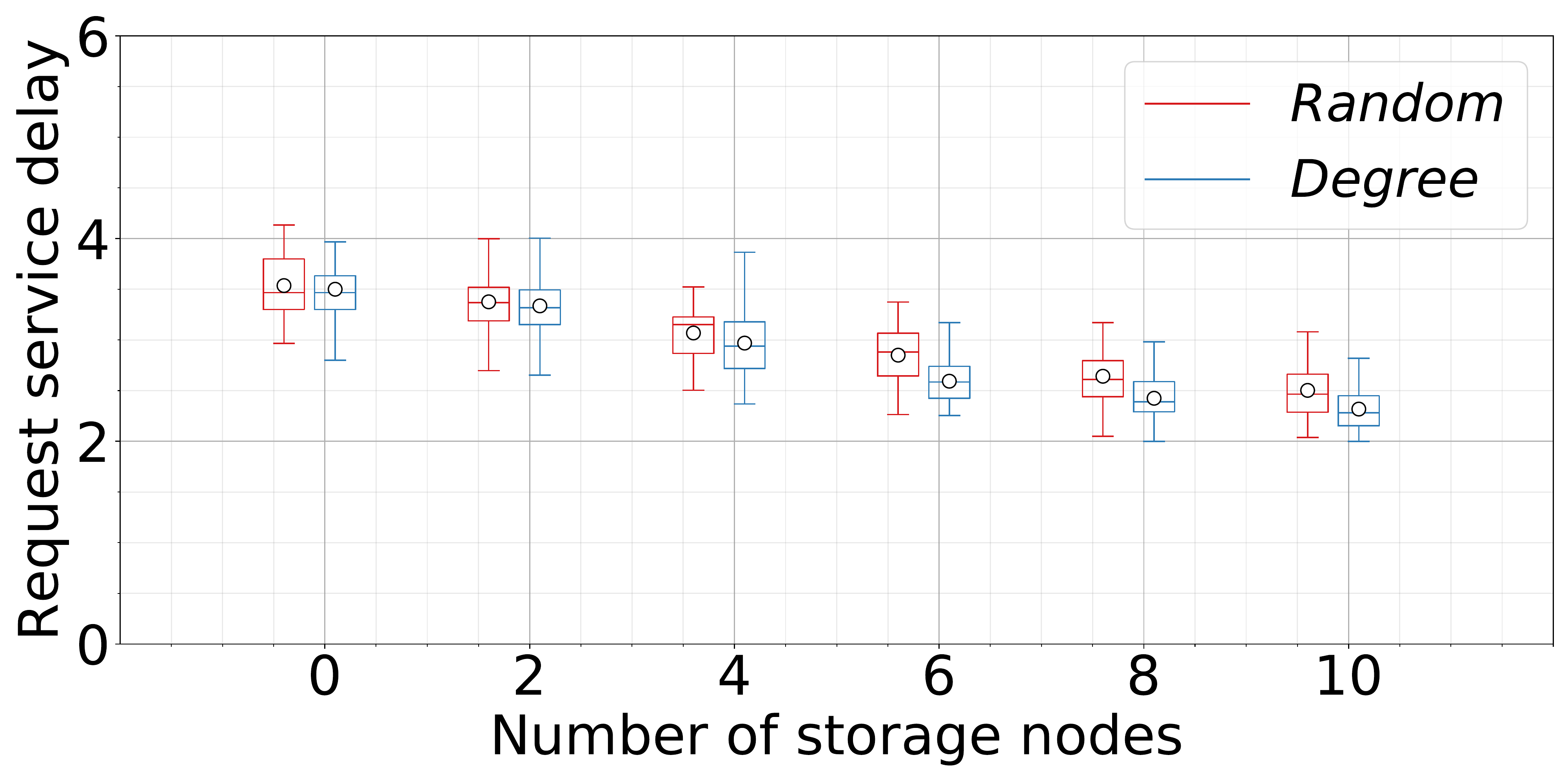}  
    \vspace{-0.06in}
    \caption{ATT}
    \label{fig:swapping_in_att2}
\end{subfigure}
\begin{subfigure}{.24\textwidth}
    \centering
    \includegraphics[width=.95\linewidth]{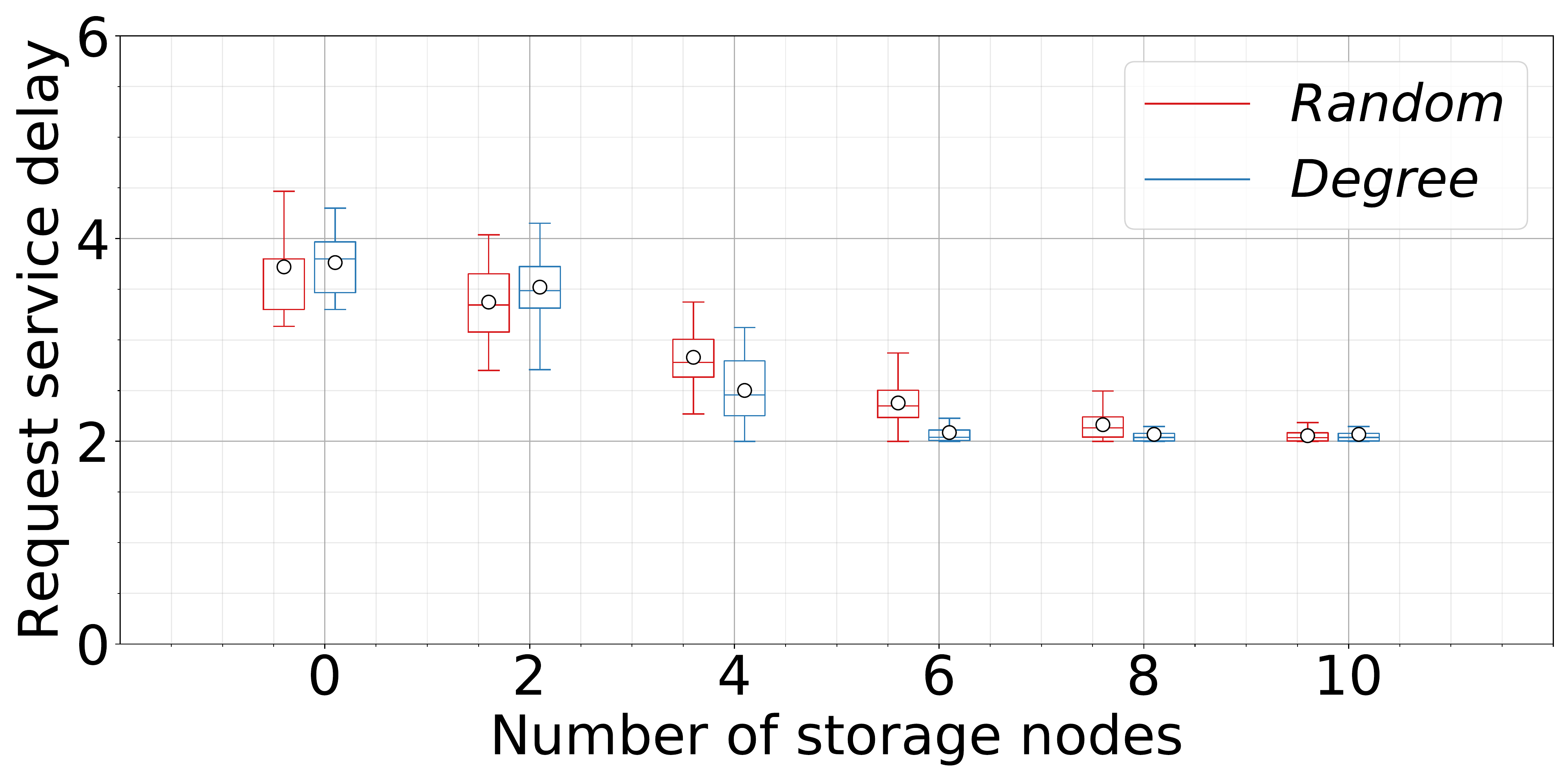}  
    \vspace{-0.06in}
    \caption{Abilene}
    \label{fig:swapping_in_abilene2}
\end{subfigure}
\begin{subfigure}{.24\textwidth}
    \centering
    \includegraphics[width=.95\linewidth]{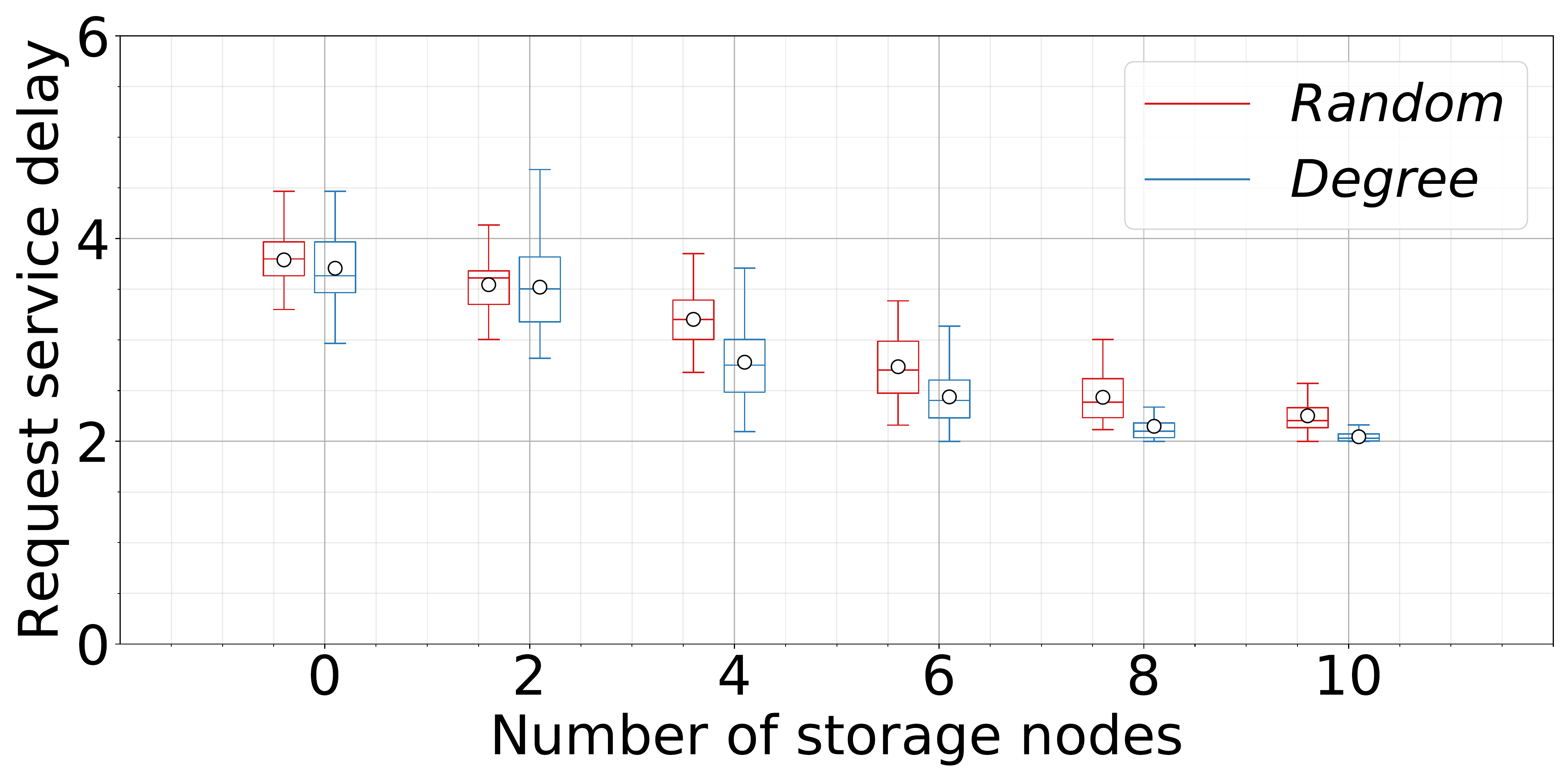}  
    \vspace{-0.06in}
    \caption{IBM}
    \label{fig:swapping_in_IBM}
\end{subfigure}
\begin{subfigure}{.24\textwidth}
    \centering
    \includegraphics[width=.95\linewidth]{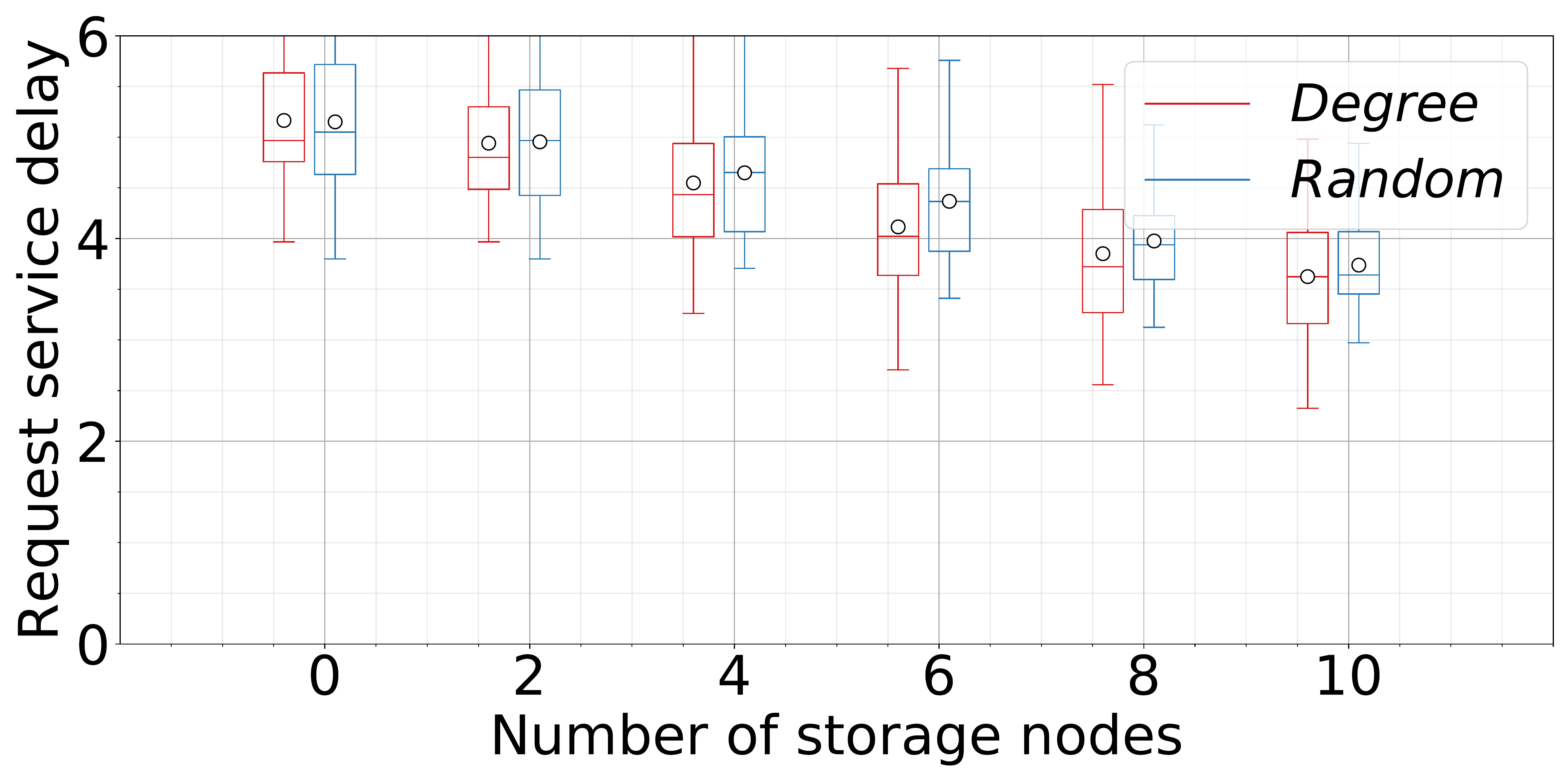}  
    \vspace{-0.06in}
    \caption{SURFnet}
    \label{fig:swapping_in_surfnet2}
\end{subfigure}

\begin{subfigure}{.24\textwidth}
    \centering
    \includegraphics[width=.95\linewidth]{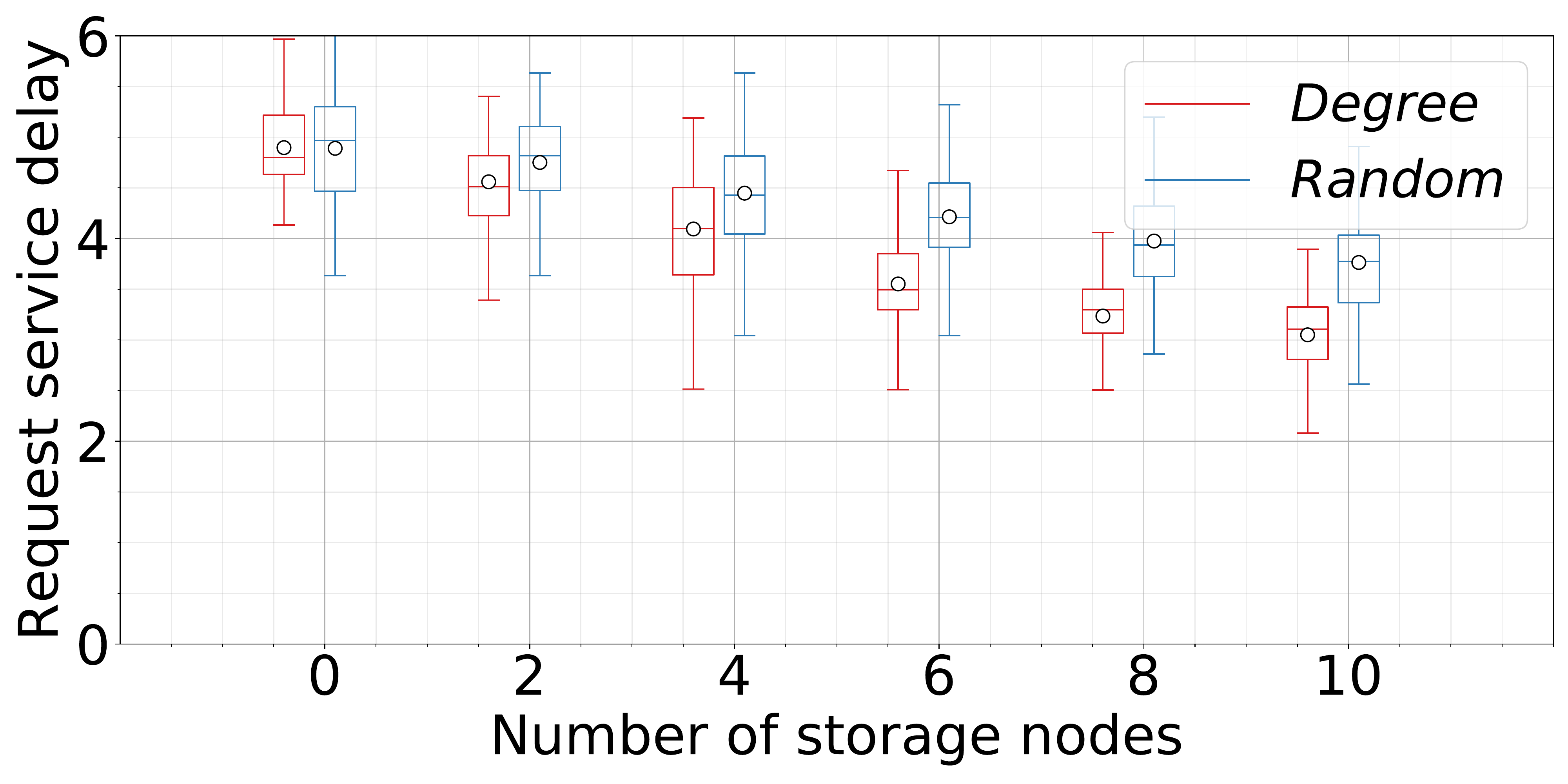}  
    \vspace{-0.06in}
    \caption{$G(50,0.05)$}
    \label{fig:swapping_in_random22}
\end{subfigure}
\begin{subfigure}{.24\textwidth}
    \centering
    \includegraphics[width=.95\linewidth]{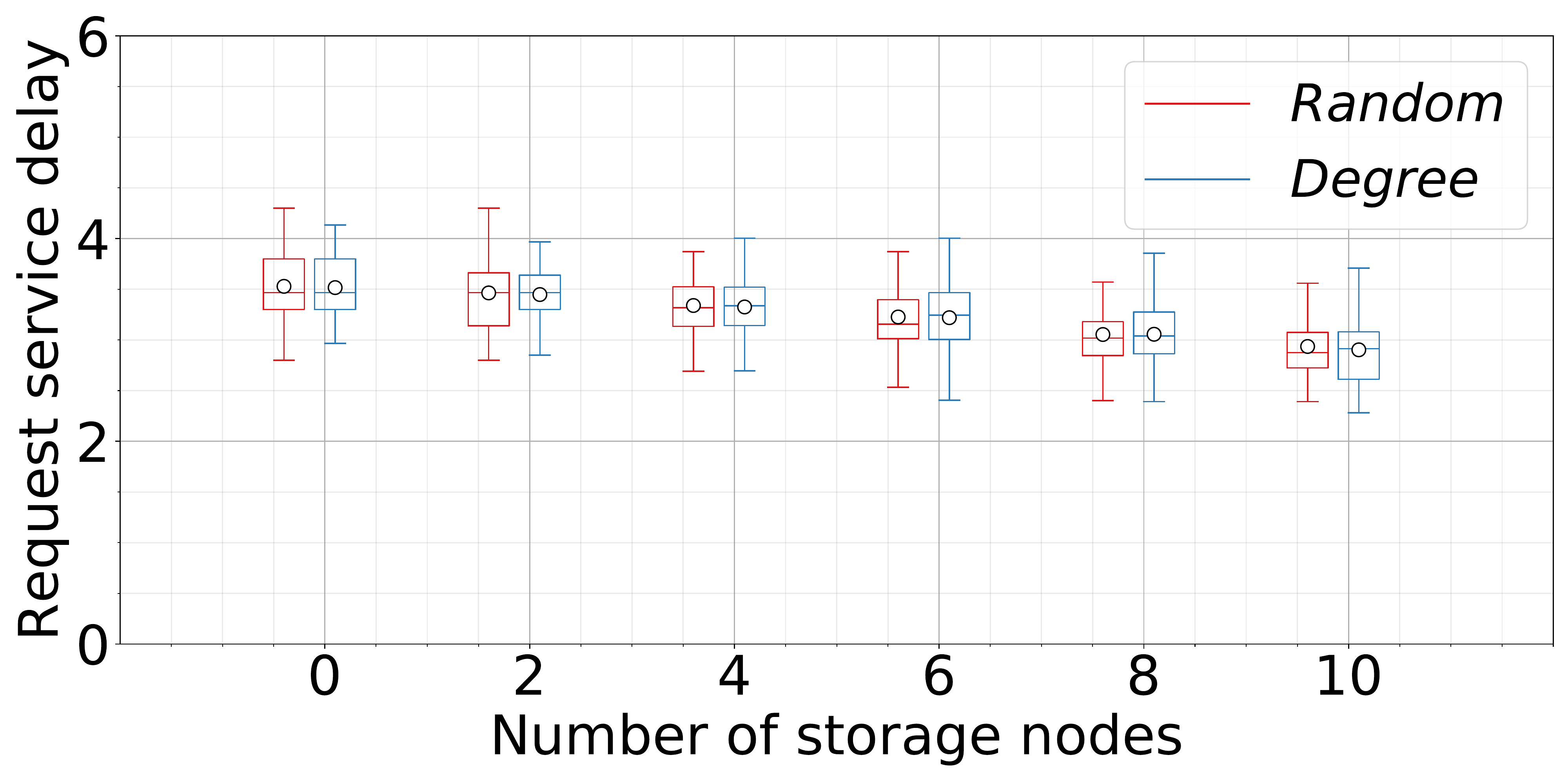} 
    \vspace{-0.06in}
    \caption{$G(50,0.1)$}
    \label{fig:swapping_in_random12}
\end{subfigure}
\begin{subfigure}{.24\textwidth}
    \centering
    \includegraphics[width=.95\linewidth]{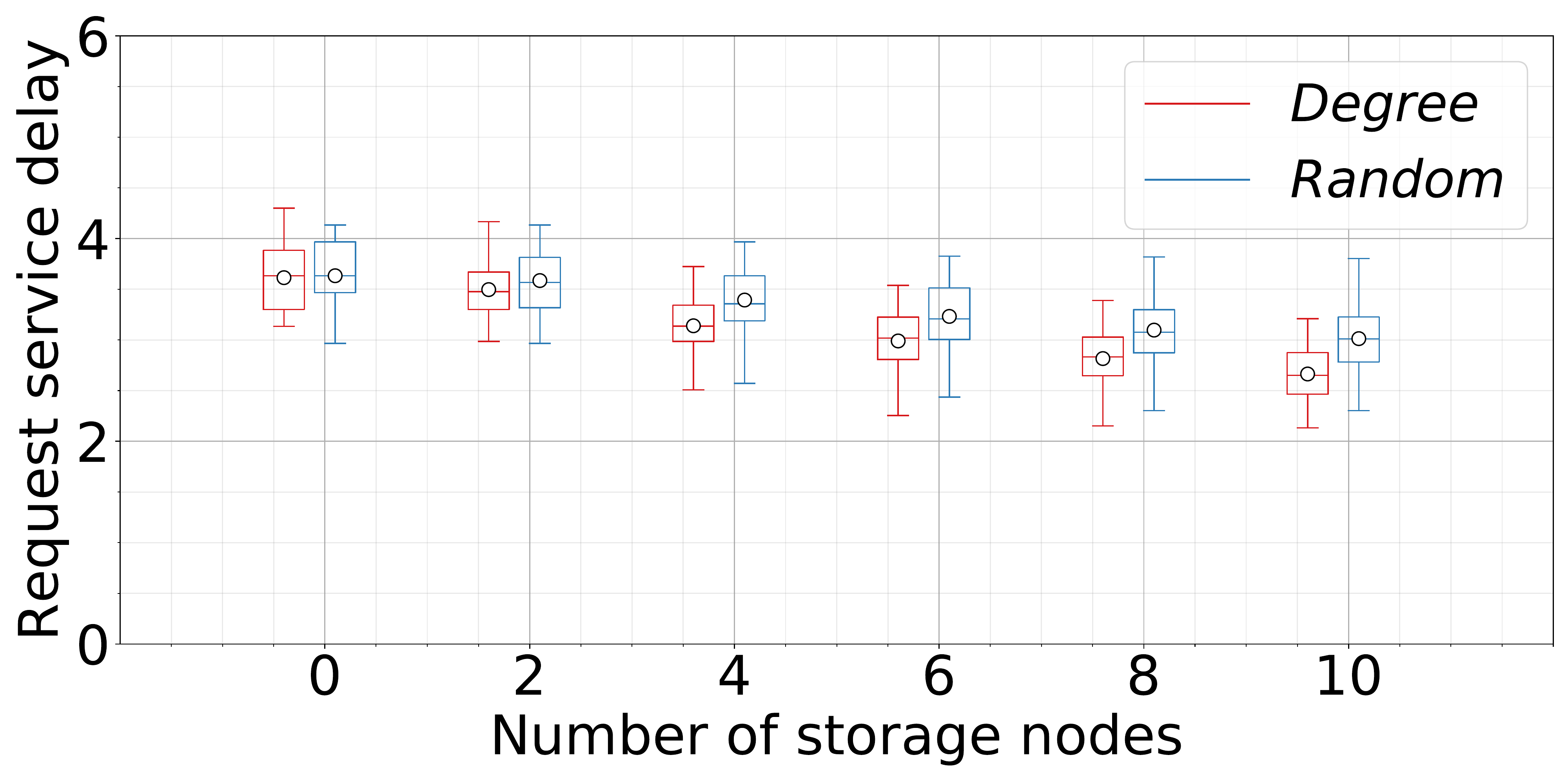}  
    \vspace{-0.06in}
    \caption{$PA(50,2)$}
    \label{fig:swapping_in_random32}
\end{subfigure}
\begin{subfigure}{.24\textwidth}
    \centering
    \includegraphics[width=.95\linewidth]{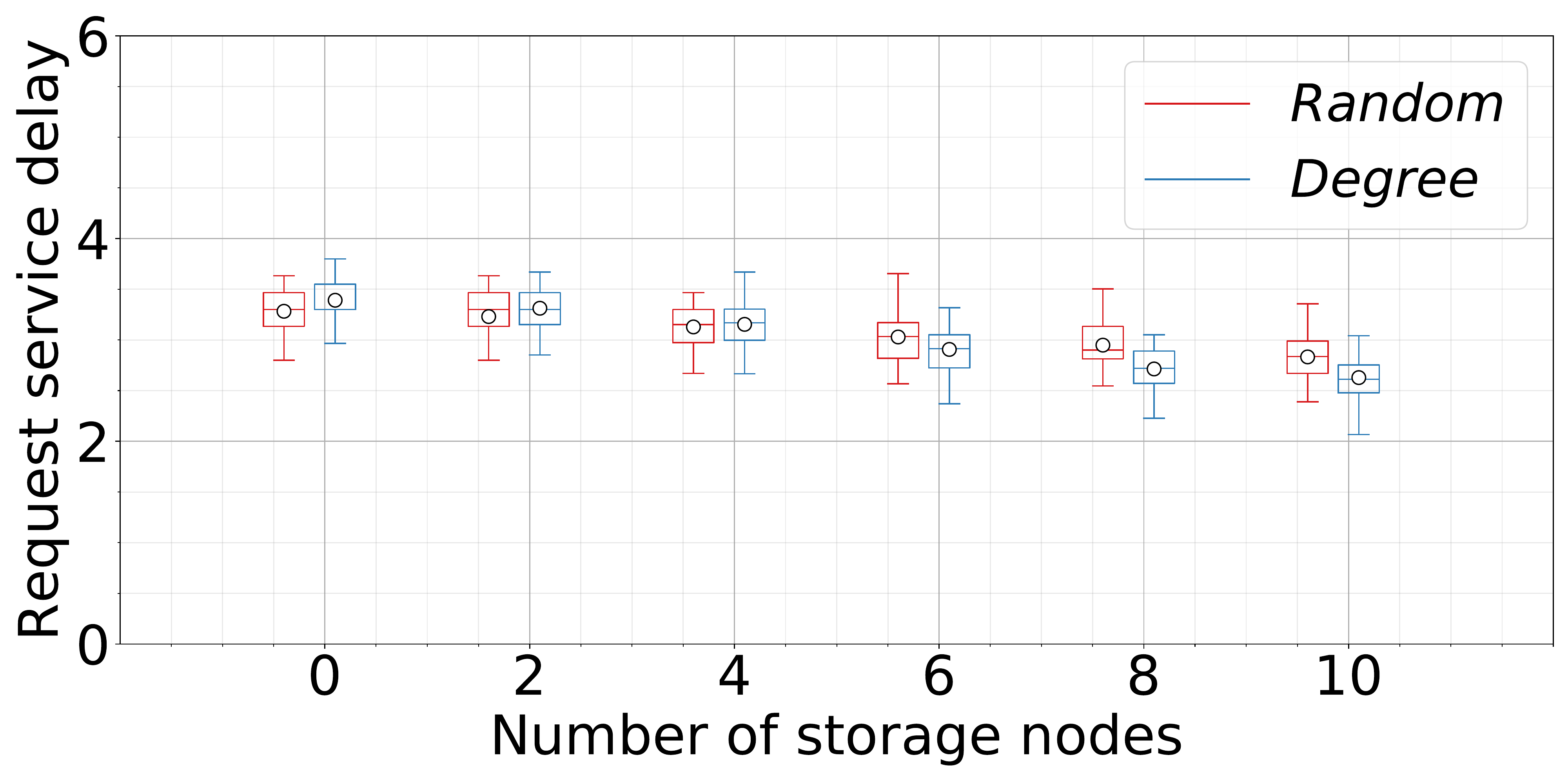}  
    \vspace{-0.06in}
    \caption{$PA(50,3)$}
    \label{fig:swapping_in_random222}
\end{subfigure}

\vspace{-0.06in}
\caption{Minimizing entanglement service delay by reducing entanglement swapping using storage servers in real and random topologies. Fidelity threshold is 0.8, $h\geq|T|$,$B_s=12000$. \label{fig:swapping}}
\end{figure*}

Figure \ref{fig:swapping} shows a box plot of the request service delays (number of entanglement swaps) in our real and random topologies as a function of number of storage nodes in the network. As expected, serving requests from storage servers reduces delays. SURFnet has the largest service delays (sub-figure \ref{fig:swapping_in_surfnet2}). This topology has the largest network diameter among all topologies (Table \ref{table:topology_charecteristics}). With $6$ or more storage nodes in the network in most topologies, we only need $2$ entanglement swap to serve the request. This can happen when each user in the user pair only needs to be connected to one storage server and $2$ entanglement swaps are required to establish the end-to-end connection.

%% file: relatedworks-new.tex
\section{Related Work}\label{sec:related}
In this Section, we discuss the state-of-the-art related to classical overlays and entanglement distribution in quantum networks.
\subsection{Classical Overlay Networks}
Broadly, almost all overlay networks designed for classical internet-based services fall into the following three categories: caching overlay \cite{Dilley1998}, routing overlay \cite{Andersen01}, \cite{Andreev03} and security overlay \cite{Keromytis02}. We refer the interested reader to \cite{Sitaraman2014} for a comprehensive discussion on different classical overlay network architectures. Inspired by classical overlays, in this work, we propose an overlay network architecture to efficiently distribute quantum entanglements between end users by utilizing an underlying quantum network. Our proposed architecture resembles more to a classical routing overlay in spirit where multiple overlay paths are constructed between same end users for routing communication.

\subsection{Entanglement Distribution in Quantum Networks}
A large and growing body of literature in quantum routing have investigated the problem of efficiently distributing long distance entanglements between end users using a network of quantum nodes. For example, Van Meter et al. \cite{VanMeter09} considered entanglement distribution in a linear quantum network. Pant et al. \cite{pant2019routing} presented a multi-path routing protocol to distribute entanglement in a grid network. Van Meter et al. \cite{VanMeter2013} explored several link cost metrics and used Dijkstra’s algorithm to compute the optimal routing path in a generic quantum network. Shi et al. \cite{Shi20} developed a nonlinear path cost based routing algorithm for a generic quantum network. Authors in \cite{victora2020purification},  \cite{zhao2022e2e} applied entanglement purification and developed fidelity aware routing protocols for general quantum networks.

In a separate line of work \cite{Schoute2016}, \cite{Chakraborty2019}, authors considered the notion of virtual links in a quantum network by assuming pre-shared entanglement among non-adjacent nodes and developed routing protocols. We outline the main differences between those approaches and our work. \cite{Schoute2016} focused on specific network topologies such as ring and sphere networks while we derive results for a general quantum network. In \cite{Schoute2016} and \cite{Chakraborty2019}, authors assumed that the virtual links were created in the background and were not concerned about its creation process. In our model we use the underlay quantum network to create multiple entanglements that can be stored in overlay storage nodes. Also, we consider the effect of end-to-end entanglement purification which is ignored in these work.

%% file: conclusion.tex
\section{Conclusion and Future Work}
\label{sec:concl}
In this paper, we presented the design, formal analysis, and evaluation of QONs. Although the concept of overlay networks is a widely embraced direction in classical networks, to our knowledge, this is the first work to present a full-fledged design and problem formulation of QONs. Our underlying contributions include the problem formulation for resource allocation in QONs with different performance objectives and demonstrating the potential for significantly increasing the entanglement generation rate and handling end-user entanglement demand spikes. Below we discuss some interesting open problems and challenges associated with QONs.
\begin{itemize} 
    \item {\bf Joint storage node placement and performance optimization:} In this work, we assumed the locations of storage nodes were given as an input to the optimization problem. However, one can solve the joint problem of placing the storage nodes and maximizing a certain performance objective.
   
    \item {\bf Overlay Network Topology:} We assumed the overlay storage network topology to be a complete graph. It would be interesting to explore the effect of other topologies on overlay performance.

    \item {\bf Link-level Purification:} The focus of this work has been on an architecture where purification is performed only at the end users. One can also consider an alternate architecture, where both link-level and end-to-end purification are performed at the intermediate nodes and end users respectively.

\end{itemize}

\section{Acknowledgements}

This research was supported by the NSF Engineering
Research Center for Quantum Networks (CQN), awarded under cooperative agreement number 1941583 and by NSF grant CNS-1955834. Any opinions, findings, conclusions, or recommendations expressed in this material are those of the authors and do not necessarily reflect the views of the National Science Foundation.

%% file: appendix.tex
\section{Appendix}

\subsection{$1\leq h\leq|T|$ case}
\label{sec:multi_time_interval_case}

In this section, we explain the case that EPR pairs can be stored at storage servers for multi-time-intervals (for the case $1\leq h\leq|T|$). We add one more dimension $h$ for the age of delivered or consumed EPR pairs for purification to variables $w^k_{p,t}$ and $u^j_{p,t}$. In our formulation, we use function $\pi$ to indicate different scheme in selecting EPR pairs from the storage. Possible candidate polices are as (1)\textit{ oldest-first scheme:} the oldest EPR pairs that have been stored at storage server would be used first, (2) \textit{newest-first scheme:} use the freshest EPR pairs first, and (3) \textit{randomly selected scheme:} EPR pairs would randomly be selected. The problem formulation would be as follows. Note that we will sum constraints \eqref{cons:inventory_usage}, 
\eqref{cons:demand_constraint}, \eqref{cons:edge_constraint}, \eqref{cons:storage_capacity_constraint}, \eqref{cons:variable_constraint}, \eqref{cons:storage_variable_constraint} on dimension $h$ as well. We have skipped this for brevity.

\begin{align} 
\max_{w^k_{p,t}}  & \quad 1 
\label{optimization:multi_max_average_fidelity}
\nonumber
\\
\text{subject to}\quad
& \quad \forall{t \in T:} \nonumber \\
\quad & u_{p_s,t}^{j,h} = \begin{cases}
0, \quad \quad \text{ if } \quad h>t\\
\\
u_{p_s,t-1}^{j,h-1} - \\ \quad \quad \pi(\sum_{{k \in \{K_t\cup \{J-j\}\}}\& {p \in P_{S}^{k}}|p_s \in f(p)}\\ \nonumber \quad w_{p,t-1}^{k,h-1} g(F_p,F^k_{t-1}))\Delta_{t-1},\\ \quad \quad \text{ if }  h\neq t,\\
w_{p_s,t-1}^{j,t-1}, \text{ if } h==t
\end{cases}
\nonumber
\\
\quad & \text{and constraints}  \quad 
\nonumber
\eqref{cons:inventory_usage}, 
\eqref{cons:demand_constraint}, \eqref{cons:edge_constraint}, \eqref{cons:storage_capacity_constraint}, \eqref{cons:variable_constraint}, \eqref{cons:storage_variable_constraint}
\end{align}

%% file: main.bbl
\begin{thebibliography}{10}

\bibitem{Sitaraman2014}
{Overlay Networks: An Akamai Perspective}.
\newblock {\em Advanced Content Delivery, Streaming, and Cloud Services}, pages
  305--328, 2014.

\bibitem{Abilene}
Abilene.
\newblock {Yin Zhang’s Abilene.. [Online].}
\newblock \url{http://www.cs.utexas.edu/yzhang/research/AbileneTM/}, 2020.
\newblock [Online; accessed 2-Nov-2021].

\bibitem{Andersen01}
David Andersen, Hari Balakrishnan, Frans Kaashoek, and Robert Morris.
\newblock Resilient overlay networks.
\newblock {\em SIGOPS Oper. Syst. Rev.}, 35(5):131–145, oct 2001.

\bibitem{Andreev03}
Konstantin Andreev, Bruce~M. Maggs, Adam Meyerson, and Ramesh~K. Sitaraman.
\newblock Designing overlay multicast networks for streaming.
\newblock In {\em Proceedings of the Fifteenth Annual ACM Symposium on Parallel
  Algorithms and Architectures}, page 149–158, New York, NY, USA, 2003.
  Association for Computing Machinery.

\bibitem{barabasi1999emergence}
Albert-L{\'a}szl{\'o} Barab{\'a}si and R{\'e}ka Albert.
\newblock Emergence of scaling in random networks.
\newblock {\em science}, 286(5439):509--512, 1999.

\bibitem{bennett2020quantum}
Charles~H Bennett and Gilles Brassard.
\newblock Quantum cryptography: Public key distribution and coin tossing.
\newblock {\em arXiv preprint arXiv:2003.06557}, 2020.

\bibitem{briegel1998quantum}
H-J Briegel, Wolfgang D{\"u}r, Juan~I Cirac, and Peter Zoller.
\newblock Quantum repeaters: the role of imperfect local operations in quantum
  communication.
\newblock {\em Physical Review Letters}, 81(26):5932, 1998.

\bibitem{chakraborty2020entanglement}
Kaushik Chakraborty, David Elkouss, Bruno Rijsman, and Stephanie Wehner.
\newblock Entanglement distribution in a quantum network: A multicommodity
  flow-based approach.
\newblock {\em IEEE Transactions on Quantum Engineering}, 1:1--21, 2020.

\bibitem{Chakraborty2019}
Kaushik Chakraborty, Filip Rozpedek, Axel Dahlberg, and Stephanie Wehner.
\newblock {Distributed Routing in a Quantum Internet}.
\newblock 2019.

\bibitem{cirac1999distributed}
J~Ignacio Cirac, AK~Ekert, Susana~F Huelga, and Chiara Macchiavello.
\newblock Distributed quantum computation over noisy channels.
\newblock {\em Physical Review A}, 59(6):4249, 1999.

\bibitem{dahlberg2019link}
Axel Dahlberg, Matthew Skrzypczyk, Tim Coopmans, Leon Wubben, Filip
  Rozp{\k{e}}dek, Matteo Pompili, Arian Stolk, Przemys{\l}aw Pawe{\l}czak,
  Robert Knegjens, Julio de~Oliveira~Filho, et~al.
\newblock A link layer protocol for quantum networks.
\newblock In {\em Proceedings of the ACM Special Interest Group on Data
  Communication}, pages 159--173. 2019.

\bibitem{d2001using}
G~Mauro D'Ariano, Paoloplacido~Lo Presti, and Matteo~GA Paris.
\newblock Using entanglement improves the precision of quantum measurements.
\newblock {\em Physical review letters}, 87(27):270404, 2001.

\bibitem{Deutsch96}
David Deutsch, Artur Ekert, Richard Jozsa, Chiara Macchiavello, Sandu Popescu,
  and Anna Sanpera.
\newblock Quantum privacy amplification and the security of quantum
  cryptography over noisy channels.
\newblock {\em Phys. Rev. Lett.}, 77:2818--2821, Sep 1996.

\bibitem{Dilley1998}
John Dilley, Bruce~M Maggs, Jay Parikh, Harald Prokop, Ramesh Sitaraman, and
  Bill Weihl.
\newblock {Globally distributed content delivery}.
\newblock 10 1998.

\bibitem{dudin2013light}
YO~Dudin, L~Li, and A~Kuzmich.
\newblock Light storage on the time scale of a minute.
\newblock {\em Physical Review A}, 87(3):031801, 2013.

\bibitem{Dur1999}
W.~D{\"{u}}r, H.~J. Briegel, J.~I. Cirac, and P.~Zoller.
\newblock {Quantum repeaters based on entanglement purification}.
\newblock {\em Physical Review A - Atomic, Molecular, and Optical Physics},
  59(1):169--181, 1999.

\bibitem{giovannetti2004quantum}
Vittorio Giovannetti, Seth Lloyd, and Lorenzo Maccone.
\newblock Quantum-enhanced measurements: beating the standard quantum limit.
\newblock {\em Science}, 306(5700):1330--1336, 2004.

\bibitem{SciPyProceedings_11}
Aric~A. Hagberg, Daniel~A. Schult, and Pieter~J. Swart.
\newblock Exploring network structure, dynamics, and function using networkx.
\newblock In Ga\"el Varoquaux, Travis Vaught, and Jarrod Millman, editors, {\em
  Proceedings of the 7th Python in Science Conference}, pages 11 -- 15,
  Pasadena, CA USA, 2008.

\bibitem{heinze2013stopped}
Georg Heinze, Christian Hubrich, and Thomas Halfmann.
\newblock Stopped light and image storage by electromagnetically induced
  transparency up to the regime of one minute.
\newblock {\em Physical review letters}, 111(3):033601, 2013.

\bibitem{heshami2016quantum}
Khabat Heshami, Duncan~G England, Peter~C Humphreys, Philip~J Bustard, Victor~M
  Acosta, Joshua Nunn, and Benjamin~J Sussman.
\newblock Quantum memories: emerging applications and recent advances.
\newblock {\em Journal of modern optics}, 63(20):2005--2028, 2016.

\bibitem{cplex}
IBM.
\newblock {IBM [Online].}
\newblock \url{https://www.ibm.com/academic}, 2022.
\newblock [Online; accessed 2-Jun-2022].

\bibitem{Keromytis02}
Angelos~D. Keromytis, Vishal Misra, and Dan Rubenstein.
\newblock Sos: Secure overlay services.
\newblock SIGCOMM '02, page 61–72, New York, NY, USA, 2002. Association for
  Computing Machinery.

\bibitem{kimble2008quantum}
H~Jeff Kimble.
\newblock The quantum internet.
\newblock {\em Nature}, 453(7198):1023--1030, 2008.

\bibitem{knight2011internet}
Simon Knight, Hung~X Nguyen, Nickolas Falkner, Rhys Bowden, and Matthew
  Roughan.
\newblock The internet topology zoo.
\newblock {\em IEEE Journal on Selected Areas in Communications},
  29(9):1765--1775, 2011.

\bibitem{komar2014quantum}
Peter Komar, Eric~M Kessler, Michael Bishof, Liang Jiang, Anders~S S{\o}rensen,
  Jun Ye, and Mikhail~D Lukin.
\newblock A quantum network of clocks.
\newblock {\em Nature Physics}, 10(8):582--587, 2014.

\bibitem{lloyd2004infrastructure}
Seth Lloyd, Jeffrey~H Shapiro, Franco~NC Wong, Prem Kumar, Selim~M Shahriar,
  and Horace~P Yuen.
\newblock Infrastructure for the quantum internet.
\newblock {\em ACM SIGCOMM Computer Communication Review}, 34(5):9--20, 2004.

\bibitem{longdell2005stopped}
Jevon~J Longdell, Elliot Fraval, Matthew~J Sellars, and Neil~B Manson.
\newblock Stopped light with storage times greater than one second using
  electromagnetically induced transparency in a solid.
\newblock {\em Physical review letters}, 95(6):063601, 2005.

\bibitem{ma2021one}
Yu~Ma, You-Zhi Ma, Zong-Quan Zhou, Chuan-Feng Li, and Guang-Can Guo.
\newblock One-hour coherent optical storage in an atomic frequency comb memory.
\newblock {\em Nature communications}, 12(1):1--6, 2021.

\bibitem{mattle1996dense}
Klaus Mattle, Harald Weinfurter, Paul~G Kwiat, and Anton Zeilinger.
\newblock Dense coding in experimental quantum communication.
\newblock {\em Physical Review Letters}, 76(25):4656, 1996.

\bibitem{pant2019routing}
Mihir Pant, Hari Krovi, Don Towsley, Leandros Tassiulas, Liang Jiang, Prithwish
  Basu, Dirk Englund, and Saikat Guha.
\newblock Routing entanglement in the quantum internet.
\newblock {\em npj Quantum Information}, 5(1):1--9, 2019.

\bibitem{peev2009secoqc}
Momtchil Peev, Christoph Pacher, Romain All{\'e}aume, Claudio Barreiro, Jan
  Bouda, W~Boxleitner, Thierry Debuisschert, Eleni Diamanti, Mehrdad Dianati,
  JF~Dynes, et~al.
\newblock The secoqc quantum key distribution network in vienna.
\newblock {\em New Journal of Physics}, 11(7):075001, 2009.

\bibitem{tmgen}
Python.
\newblock {TMgen API[Online].}
\newblock
  \url{https://tmgen.readthedocs.io/en/latest/api.html#tmgen.models.spike_tm},
  2022.
\newblock [Online; accessed 2-Jun-2022].

\bibitem{sangouard2011quantum}
Nicolas Sangouard, Christoph Simon, Hugues De~Riedmatten, and Nicolas Gisin.
\newblock Quantum repeaters based on atomic ensembles and linear optics.
\newblock {\em Reviews of Modern Physics}, 83(1):33, 2011.

\bibitem{Schoute2016}
Eddie Schoute, Laura Mancinska, Tanvirul Islam, Iordanis Kerenidis, and
  Stephanie Wehner.
\newblock {Shortcuts to quantum network routing}.
\newblock pages 1--45, 2016.

\bibitem{Shi20}
Shouqian Shi and Chen Qian.
\newblock Concurrent entanglement routing for quantum networks: Model and
  designs.
\newblock SIGCOMM '20, page 62–75, New York, NY, USA, 2020. Association for
  Computing Machinery.

\bibitem{Teavar_source_code}
Teavar source code.
\newblock {Traffic Engineering Applying Value at Risk tool source code
  [Online].}
\newblock \url{https://github.com/manyaghobadi/teavar}, 2020.
\newblock [Online; accessed 2-Nov-2020].

\bibitem{stucki2011long}
Damien Stucki, Matthieu Legre, Francois Buntschu, B~Clausen, Nadine Felber,
  Nicolas Gisin, Luca Henzen, Pascal Junod, G{\'e}rald Litzistorf, Patrick
  Monbaron, et~al.
\newblock Long-term performance of the swissquantum quantum key distribution
  network in a field environment.
\newblock {\em New Journal of Physics}, 13(12):123001, 2011.

\bibitem{VanMeter09}
Rodney Van~Meter, Thaddeus~D. Ladd, W.~J. Munro, and Kae Nemoto.
\newblock System design for a long-line quantum repeater.
\newblock {\em IEEE/ACM Trans. Netw.}, 17(3):1002–1013, jun 2009.

\bibitem{VanMeter2013}
Rodney {Van Meter}, Takahiko Satoh, Thaddeus~D Ladd, William~J Munro, and Kae
  Nemoto.
\newblock {Path selection for quantum repeater networks}.
\newblock {\em Networking Science}, 3(1):82--95, 2013.

\bibitem{victora2020purification}
Michelle Victora, Stefan Krastanov, Alexander~Sanchez de~la Cerda, Steven
  Willis, and Prineha Narang.
\newblock Purification and entanglement routing on quantum networks.
\newblock {\em arXiv preprint arXiv:2011.11644}, 2020.

\bibitem{wang2014field}
Shuang Wang, Wei Chen, Zhen-Qiang Yin, Hong-Wei Li, De-Yong He, Yu-Hu Li, Zheng
  Zhou, Xiao-Tian Song, Fang-Yi Li, Dong Wang, et~al.
\newblock Field and long-term demonstration of a wide area quantum key
  distribution network.
\newblock {\em Optics express}, 22(18):21739--21756, 2014.

\bibitem{werner1989quantum}
Reinhard~F Werner.
\newblock Quantum states with einstein-podolsky-rosen correlations admitting a
  hidden-variable model.
\newblock {\em Physical Review A}, 40(8):4277, 1989.

\bibitem{zhao2021redundant}
Yangming Zhao and Chunming Qiao.
\newblock Redundant entanglement provisioning and selection for throughput
  maximization in quantum networks.
\newblock In {\em IEEE INFOCOM 2021-IEEE Conference on Computer
  Communications}, pages 1--10. IEEE, 2021.

\bibitem{zhao2022e2e}
Yangming Zhao, Gongming Zhao, and Chunming Qiao.
\newblock E2e fidelity aware routing and purification for throughput
  maximization in quantum networks.
\newblock In {\em Proceedings of the IEEE INFOCOM}, 2022.

\end{thebibliography}
